\documentclass[10pt,journal,compsoc]{IEEEtran}

\usepackage{amsmath,amssymb,amsfonts}
\usepackage{graphicx}
\usepackage{amsthm}
\usepackage{multirow}
\usepackage{bibentry}
\usepackage{listings} 
\usepackage{float}
\usepackage{color}
\usepackage{url}

\usepackage{ragged2e}

\usepackage{algorithm,algpseudocode}
\usepackage{amsmath}
\usepackage{amsfonts}
\usepackage{amssymb}

\usepackage{tabularx}
\usepackage{diagbox}

\usepackage[caption=false,font=normalsize,labelfont=sf,textfont=sf]{subfig}

\def\be{ \begin{eqnarray} }
\def\ee{ \end{eqnarray} }

\ifCLASSOPTIONcompsoc
  \usepackage[nocompress]{cite}
\else
  \usepackage{cite}
\fi

\ifCLASSINFOpdf
\else
\fi

\begin{document}

\title{Accuracy-Aware Cooperative Sensing and Computing for Connected Autonomous Vehicles}

\author{Xuehan Ye, Kaige Qu, \IEEEmembership{Member, IEEE}, Weihua Zhuang, \IEEEmembership{Fellow, IEEE}, \\and Xuemin (Sherman) Shen, \IEEEmembership{Fellow, IEEE}%
\thanks{Manuscript Received 11 July 2023; revised 20 November 2023; accepted 12
December 2023. Recommended for acceptance by M. Chen. This work was supported by the Natural Sciences and Engineering Research Council (NSERC) of Canada. (Corresponding author: Kaige Qu.)}
\thanks{Xuehan Ye was with the Department of Electrical and Computer Engineering, University of Waterloo, Waterloo, ON, Canada, N2L 3G1 (email: x49ye@uwaterloo.ca).}
\thanks{Kaige Qu, Weihua Zhuang, and Xuemin (Sherman) Shen are with the Department of Electrical and Computer Engineering, University of Waterloo, Waterloo, ON, Canada, N2L 3G1 (emails: \{k2qu, wzhuang, sshen\}@uwaterloo.ca).}
\thanks{Color versions of one or more figures in this article are available at https://doi.org/10.1109/TMC.2023.3343709.}
\thanks{Digital Object Identifier 10.1109/TMC.2023.3343709}}

\IEEEtitleabstractindextext{

\begin{abstract}
\justifying
To maintain high perception performance among connected and autonomous vehicles (CAVs), in this paper, we propose an accuracy-aware and resource-efficient raw-level cooperative sensing and computing scheme among CAVs and road-side infrastructure. The scheme enables fined-grained partial raw sensing data selection, transmission, fusion, and processing in per-object granularity, by exploiting the parallelism among object classification subtasks associated with each object. A supervised learning model is trained to capture the relationship between the object classification accuracy and the data quality of selected object sensing data, facilitating accuracy-aware sensing data selection. We formulate an optimization problem for joint sensing data selection, subtask placement and resource allocation among multiple object classification subtasks, to minimize the total resource cost while satisfying the delay and accuracy requirements. A genetic algorithm based iterative solution is proposed for the optimization problem. Simulation results demonstrate the accuracy awareness and resource efficiency achieved by the proposed cooperative sensing and computing scheme, in comparison with benchmark solutions. 
\end{abstract}

\begin{IEEEkeywords}
Connected and autonomous vehicles (CAVs), environment perception, cooperative sensing, cooperative computing, supervised learning, vehicular edge computing. 
\end{IEEEkeywords}}

\maketitle

\IEEEdisplaynontitleabstractindextext

\IEEEpeerreviewmaketitle

\ifCLASSOPTIONcompsoc
\IEEEraisesectionheading{\section{Introduction}\label{sec:introduction}}
\else
\section{Introduction}
\label{sec:Introduction}
\fi

\IEEEPARstart{A}{utonomous} driving is a key use case that will reshape the future transportation systems. 
The foundation of autonomous driving is the capability for connected and autonomous vehicles (CAVs) to know their surrounding environments, specifically the locations, dimensions, and types of nearby objects, 
referred to as environment perception, based on which different autonomous driving applications can be supported, such as maneuver control and path planning~\cite{zhang2019mobile,malandrino2021edge,zhuang2019sdn,shen2021holistic,wang2018networking}. 
For environment perception, a CAV is equipped with various on-board sensors such as cameras, light detection and ranging (LiDAR) sensors, and radar sensors, to collect raw sensing data of the environment~\cite{cui2020offloading}. 
Due to line-of-sight sensing, the sensing ranges of on-board sensors are easy to be occluded by surrounding obstacles, leading to difficult detection of objects in the blind zones~\cite{li2021towards}.
Moreover,  even without occlusion, a CAV observes each object in the environment from a limited viewing angle, providing limited sensing data diversity by the on-board sensors. 
Hence, it is not reliable to fully rely on the on-board sensors for a consistent guarantee of complete and accurate environment perception.

By leveraging the emerging vehicles-to-everything (V2X) communication technologies, 
\emph{cooperative sensing} (or \emph{cooperative perception}) has been proposed to enable sharing of raw sensing data (e.g., camera images, LiDAR point clouds, radar measurements), higher-level features extracted from raw data (e.g., convolution layer output feature maps), or lightweight sensing outcomes (e.g., object detection results, alarm messages) among CAVs and infrastructures in proximity, corresponding to raw-level, feature-level, and decision-level cooperative sensing respectively~\cite{arnold2020cooperative,abdel2021vehicular,abdellatif2021active}. 
Among these cooperative sensing levels, there exists a trade-off between resource efficiency and performance enhancement in terms of perception range extension and perception accuracy improvement. 
In comparison with feature-level and decision-level cooperative sensing based on compressed/processed sensing data, raw-level cooperative sensing achieves the best perception performance by retaining the most fine-grained environmental details~\cite{chen2019cooper,jia2023mass}.  
However, due to the network resource inefficiency for transmitting and processing the large raw data, conventional broadcast-based cooperative sensing strategies for lightweight compressed/processed sensing data cannot be directly applied in raw-level cooperative sensing~\cite{ETSI,wang2020v2vnet}.

To enhance the resource efficiency while maintaining the high perception performance, existing works have investigated scalable raw-level cooperative sensing by sharing \emph{partial} raw sensing data, which allows the transmission and processing of the most relevant segments in the full raw sensing data~\cite{zhang2021emp,qiu2021autocast}. For example, a common region of interest (RoI) of multiple CAVs is partitioned to disjoint non-overlapping spatial areas, and each CAV is responsible for sharing only a segment of raw sensing data for its closest area~\cite{zhang2021emp}. 
As compared with a basic scheme in which all CAVs share the full raw sensing data, the communication and computation resources are greatly reduced without a significant perception accuracy loss. 
To further reduce the irrelevant raw sensing data, a more fine-grained raw-level cooperative sensing scheme is proposed in~\cite{qiu2021autocast}, which facilitates sharing only the partial raw sensing data of relevant objects in the scene, referred to as \emph{object sensing data}, by removing the background information. 
With the consideration that data fusion from diverse viewing angles of different CAVs enhances the perception accuracy, the sensing data for each object in the RoI are provided by multiple CAVs in~\cite{qiu2021autocast}. 
In the existing works on scalable raw-level cooperative sensing, the principles in partial raw sensing data selection are mainly for performance gain from the localization perspective, e.g., the proximity principle~\cite{zhang2021emp} or the relevance of object locations on CAVs' future trajectories~\cite{qiu2021autocast}. 
However, it is difficult to estimate the relevance of partial raw sensing data of each CAV from the perspective of improving the object classification accuracy.   
As the ground-truth object classification accuracy is unknown until all the selected partial raw sensing data from different CAVs are fused and processed by a black-box artificial intelligence (AI) model, it is challenging to predetermine what combinations of partial raw sensing data from different CAVs can achieve accuracy satisfaction. 
Therefore, it is desirable to develop an \emph{accuracy-aware} and \emph{resource-efficient} raw-level cooperative sensing scheme, which facilitates efficient selection of partial raw sensing data from different CAVs, for object classification accuracy and minimum total communication and computing resource cost during the data transmission, fusion, and processing stages. 
In this work, we focus on fine-grained and accuracy-aware partial raw sensing data selection on a per-object basis for scalable raw-level cooperative sensing, which is facilitated by per-object sensing data extraction from full raw sensing data based on lightweight data pre-processing techniques such as bounding box detection.  
Specifically, we aim to determine for each object a set of CAVs that provide the corresponding object sensing data for fusion and AI model processing. 

In addition to the accuracy-aware efficient sensing data selection issue, another research question is where to perform the data fusion and processing with improved delay performance. 
In the conventional broadcast-based cooperative sensing scheme, each CAV locally fuses and processes its own data and all the received sensing data from neighboring CAVs, which incurs large overall computing resource consumption and may lead to delay violation due to the on-board computing resource limitation. 
To satisfy the stringent perception delay requirement, existing studies focus on the infrastructure support for cooperative sensing of CAVs~\cite{zhang2021emp,zhang2021dynamic,xiao2022perception}. 
By uploading the selected partial raw sensing data from each CAV via vehicle-to-infrastructure (V2I) transmission to an edge server for computation, the more powerful edge computing resources are leveraged, the redundant computation among CAVs for identifying common objects is reduced, and each CAV can obtain the perception results with reduced latency~\cite{zhang2021emp}.  
However, to support the data fusion and processing of selected object sensing data in our fine-grained cooperative sensing scheme, purely relying on a centralized edge server is not the most resource-efficient approach. 
As the object classification can be parallelized in per-object (or per region of interest) granularity, the computing subtasks for the classification of each object can be supported with distributed computing~\cite{9712236,zhang2021elf,wang2022vabus}.
Such parallelism provides an opportunity for further delay improvement.
Thus, we consider the placement of each object classification subtask at either an edge server or one of the CAVs, to fully utilize the network-wide computing resources~\cite{liu2022rl,9808399}. 
In this manner, the CAVs not only cooperatively sense the environment but also collaborate with the edge server for computation, referred to as a \emph{cooperative sensing and computing scheme}. 
Moreover, an interesting observation is the correlation between the object sensing data selection decisions for cooperative sensing and the subtask placement decisions for cooperative computing. 
For an object classification subtask, if a CAV is selected not only to provide the corresponding object sensing data but also to support the computation of the subtask, data transmission is not required, which further enhances the communication resource efficiency and potentially reduces the delay.
Therefore, we want to develop an accuracy-aware and resource-efficient cooperative sensing and computing scheme, to jointly determine the sensing data selection and placement for all object classification subtasks and the communication and computing resource allocation for transmitting and processing the selected sensing data with delay satisfaction.  
To achieve this goal, we make the following contributions in this paper. 

\begin{itemize}

  \item 
  We propose a learning-based accuracy estimation method, which trains a deep neural network (DNN), specifically a multi-layer perception model with two hidden layers, to estimate the object classification accuracy. 
  We define a data quality indicator for any object sensing data with or without data fusion to characterize the data volume and spatial distribution, and use a bounding box detection algorithm to determine the object dimensions, both of which composite the input of the DNN;   

  \item 
  Based on the accuracy estimation learning model, we propose a cooperative sensing and computing scheme for edge-assisted CAVs, with accuracy-aware sensing data selection in per-object granularity and distributed computing among CAVs and edge server;  

  \item We formulate a joint data selection, subtask placement, and resource allocation problem, to find the optimal cooperative sensing and computing strategy among CAVs and edge server, for a minimum total computing and communication resource cost with delay and accuracy satisfaction;

  \item To solve the optimization problem, we propose a genetic algorithm based iterative solution, which iteratively updates the data selection and subtask placement decisions until convergence, based on the feasibility and optimal resource cost obtained by solving a resource allocation subproblem;  

  \item Simulation results demonstrate both accuracy improvement and resource efficiency achieved by the proposed cooperative sensing and computing scheme, in comparison with benchmark solutions. 

\end{itemize}

The remainder of this paper is organized as follows. The system model is described in Section \ref{sec:System Model}. Section \ref{sec:Joint Data Selection, Subtask Placement and Resource Allocation Problem} presents the joint data selection, subtask placement and resource allocation problem, with a solution given in Section~\ref{sec:Problem Solution}. Simulation results are discussed in Section \ref{sc:sim}, and conclusions are drawn in Section~\ref{sc:con}. 
Table I summarizes the main mathematical symbols.

\section{System Model}
\label{sec:System Model}

\begin{table}
\footnotesize
\centering
\caption{\scshape{List of important notations}}
\begin{tabular}{c  | p{0.77\columnwidth}}
\hline\hline
\multicolumn{2}{c}{\textbf{Parameters}}\tabularnewline
\hline
$\hat{a}_m$   &  Estimated object classification accuracy for subtask $m$; \tabularnewline
$\mathcal{D}_n^{(m)}$ & Object sensing data of CAV $n$ for object $m$; \tabularnewline
$f_n$    & Computing resources (in cycle/s) at computing node $n$; \tabularnewline
$\mathcal{N}$ ($\mathcal{N}^+$) &  Set of CAVs (computing nodes); \tabularnewline
$R_{n,n'}$     & Transmission rate between CAV $n$ and computing node $n'$; \tabularnewline
$t_n$     & Total computing time for all subtasks at computing node $n$; \tabularnewline
$t_{n,n'}$  & Average transmission time for all the sensing data transmitted from CAV $n$ to computing node $n'$; \tabularnewline
$\boldsymbol{Z}_n^{(m)}$  &   Data quality indicator for object sensing data $\mathcal{D}_n^{(m)}$; \tabularnewline
$\epsilon$    & Computation intensity (in cycle/point); \tabularnewline
$\mu^{(m)}$    &   Computing demand in CPU cycles of subtask $m$; \tabularnewline 
$\rho_{n,n'}$    & Total size of the sensing data transmitted from CAV $n$ to computing node $n'$; \tabularnewline
$\varphi$    & Data size (in bit) of one observation point; \tabularnewline
\hline
\multicolumn{2}{c}{\textbf{Decision variables}}\tabularnewline
\hline
$e_n^{(m)}$  &Binary variable indicating whether or not subtask $m$ is placed at computing node $n$; \tabularnewline
$s_n^{(m)}$  & Binary variable indicating whether or not object sensing data of CAV $n$ for object $m$ are selected for subtask $m$; \tabularnewline
$\alpha_n$   & Fraction of computing resource usage at computing node $n$; \tabularnewline
$\beta_{n,n'}$  & Fraction of bandwidth allocated to the communication link from CAV $n$ to computing node $n'$; \tabularnewline
$\chi_{n,n'}$     & Binary variable indicating whether or not link between CAV $n$ and computing node $n'$ is activated for data transmission; \tabularnewline
\hline\hline
\end{tabular}
\end{table}

\subsection{Edge-Assisted Autonomous Driving Scenario}

\begin{figure}
\centering
\includegraphics[width=1\linewidth]{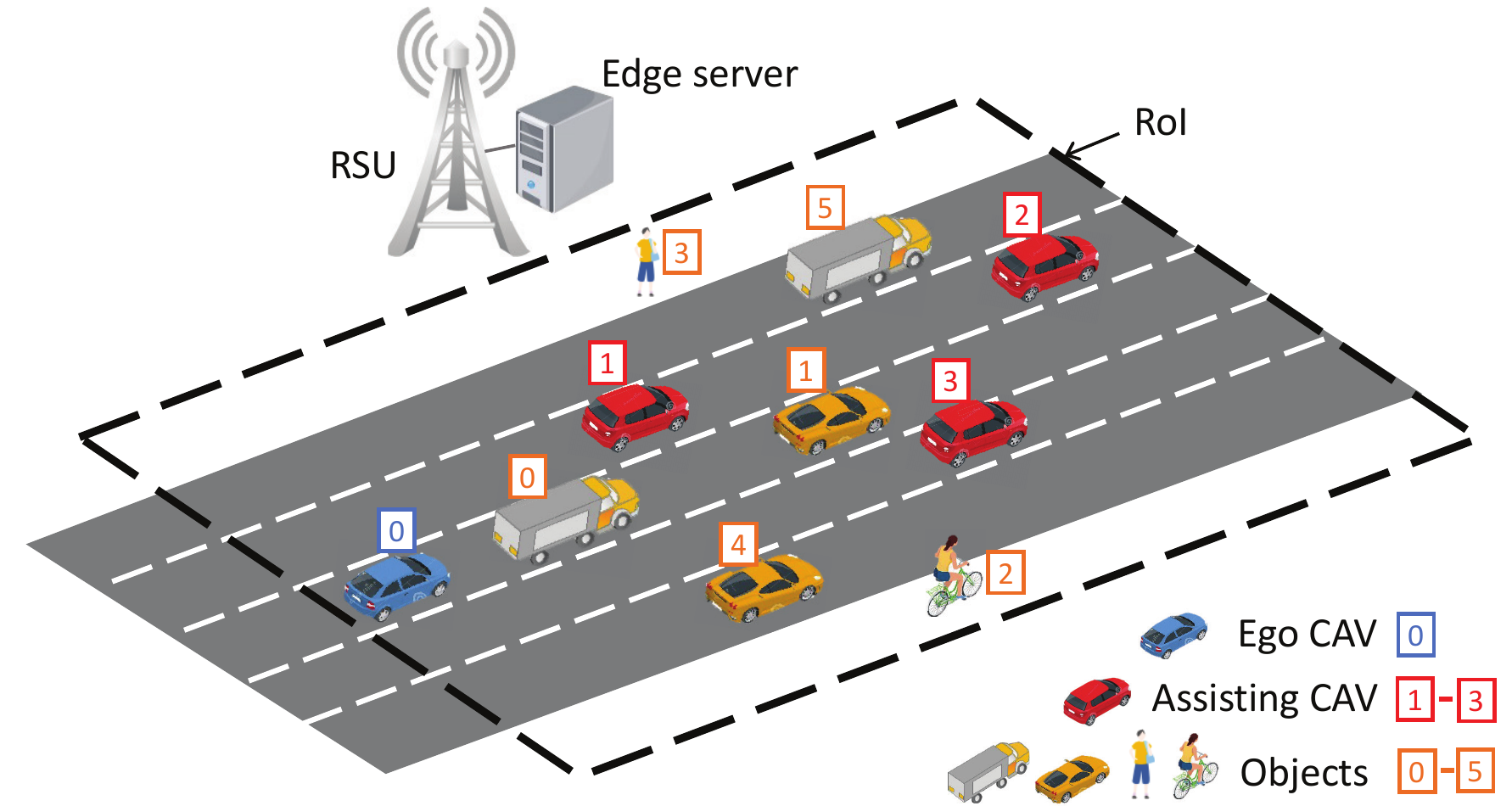}
\caption{An illustration of edge-assisted autonomous driving scenario.}
\label{fig:Network_scenario}
\end{figure}

As shown in Fig.~\ref{fig:Network_scenario}, 
we consider an edge-assisted autonomous driving scenario over a unidirectional urban road segment in the coverage of one road side unit (RSU). The RSU is co-located with an edge server, providing edge computing capability. 
Consider $N$ CAVs, including one ego CAV which initiates a perception task in an RoI and $N-1$ nearby assisting CAVs which can cooperate with the ego CAV for environment perception.
Let $\mathcal{N}=\{0,\cdots,N-1\}$ denote the set of CAVs, with $n\in\mathcal{N}$ representing the CAV index.
Specifically, CAV 0 corresponds to the ego CAV.
Let $\mathcal{N}^+ = \mathcal{N} \cup \{N\}$ denote the set of computing nodes, including all the CAVs in set $\mathcal{N}$ and the RSU which is referred to as computing node $N$. 
In Fig.~\ref{fig:Network_scenario}, we have $N=4$. 

The RoI can have a customized shape according to the application scenario, e.g., a circular or rectangular area around or in the driving direction of the ego CAV~\cite{abdel2021vehicular,lv2022edge,9874998}. 
Fig.~\ref{fig:Network_scenario} illustrates an example rectangular RoI spreading over the front area of the ego CAV. 
Let $\mathcal{M}=\{0,\cdots,M-1\}$ be the set of objects in the RoI, and let $L$ denote the number of object classes. 
Fig.~\ref{fig:Network_scenario} illustrates $M=6$ objects in the RoI and $L=4$ object classes including car, truck, pedestrian and cyclist. 
The perception task of the ego CAV is to 1) detect the existence and spatial locations of all the $M$ objects, and 2) estimate the class for each detected object $m\in\mathcal{M}$.

\subsection{Sensing Data Model}
\label{sec:Sensing Data Model}

Consider that each CAV periodically scans the environment by a $360^\circ$ LiDAR sensor mounted on the roof. 
For each scan, the LiDAR sensor of CAV $n$ generates a 3D point cloud as raw sensing data, denoted by set $\mathcal{D}_n =\left\{ \left(x_{n}^i,y_{n}^i,z_{n}^i\right) \right\}$. 
The $i$-th element of $\mathcal{D}_n$, i.e., $\left(x_{n}^i,y_{n}^i,z_{n}^i\right)$, denotes the 3D coordinates of one observation point in the environment, which can be on the surface of an object, a CAV, or on the ground.  
Consider a global coordinate system at all CAVs, as the local coordinate systems can be aligned with the global one via coordinate transformation~\cite{hu2022where2comm}.

\begin{figure}[!t]
\centering
\subfloat[]{\includegraphics[width=3.2in]{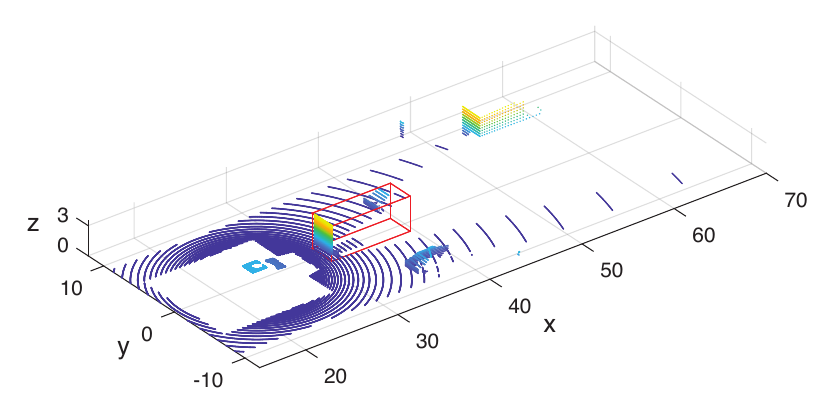}%
\label{fig:dd1}}
\hfil
\subfloat[]{\includegraphics[width=3.2in]{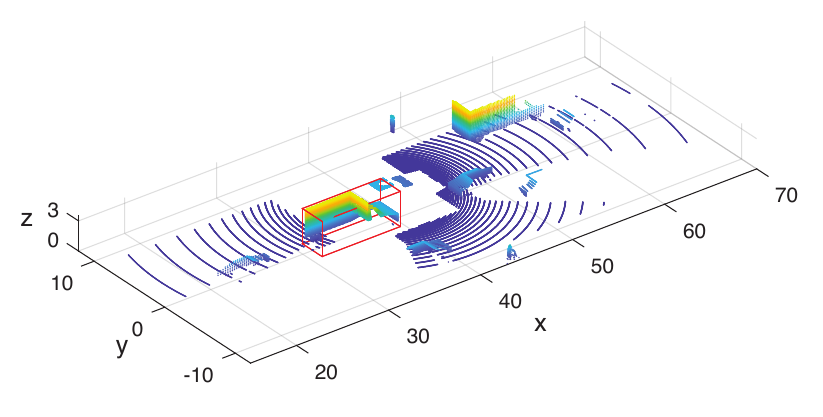}%
\label{fig:dd2}}
\caption{The simulated LiDAR 3D point clouds of two CAVs. (a) Provided by the ego CAV. (b) Provided by an assisting CAV.}
\label{fig:Pointcloud}
\end{figure}

The spatial location of object $m\in\mathcal{M}$ is indicated by a 3D cuboid bounding box containing the object, represented by a 9-tuple, $\boldsymbol{c}_m = \left(x_m,y_m,z_m,l_m^{(x)},l_m^{(y)},l_m^{(z)},\delta_m^{(x)},\delta_m^{(y)},\delta_m^{(z)}\right)$, 
where $x_m,y_m,z_m$ specify the 3D coordinates of the cuboid center,  $l_m^{(x)},l_m^{(y)},l_m^{(z)}$ specify the lengths of the cuboid along the $x$, $y$, and $z$ axes, and $\delta_m^{(x)},\delta_m^{(y)},\delta_m^{(z)}$ specify the rotation angles for the cuboid along the $x$, $y$, and $z$ axes.
Without loss of generality, we assume zero rotation angles for all the objects, and use a 6-tuple, $\boldsymbol{c}_m = \left(x_m,y_m,z_m,l_m^{(x)},l_m^{(y)},l_m^{(z)}\right)$, to represent the 3D bounding box for simplicity. 
Given the bounding box parameters, $\boldsymbol{c}_m$, for object $m$, the object sensing data for object $m$ at CAV $n$, denoted by $\mathcal{D}_n^{(m)}$, can be extracted from full raw sensing data $\mathcal{D}_n$, which include a subset of observation points located inside the bounding box for object $m$, satisfying 
$x_m - l_m^{(x)}/2 \leq x_{n}^i \leq x_m + l_m^{(x)}/2$, $y_m - l_m^{(y)}/2 \leq y_{n}^i \leq y_m + l_m^{(y)}/2$, and $z_m - l_m^{(z)}/2 \leq z_{n}^i \leq z_m + l_m^{(z)}/2$. 
If object $m$ is not viewed by CAV $n$ due to distance or occlusion, $\mathcal{D}_n^{(m)}$ is an empty set.  
Fig.~\ref{fig:Pointcloud} shows the 3D point clouds at ego CAV 0 and assisting CAV 1 in Fig.~\ref{fig:Network_scenario}, with red bounding boxes indicating the spatial location of object 0, a truck. It can be observed that the observation points of the ego CAV for the truck are concentrated at the back side, while the observation points of the assisting CAV for the truck mainly spread over the front and left sides.

\begin{figure}
    \centering
        \includegraphics[width=0.32\textwidth]{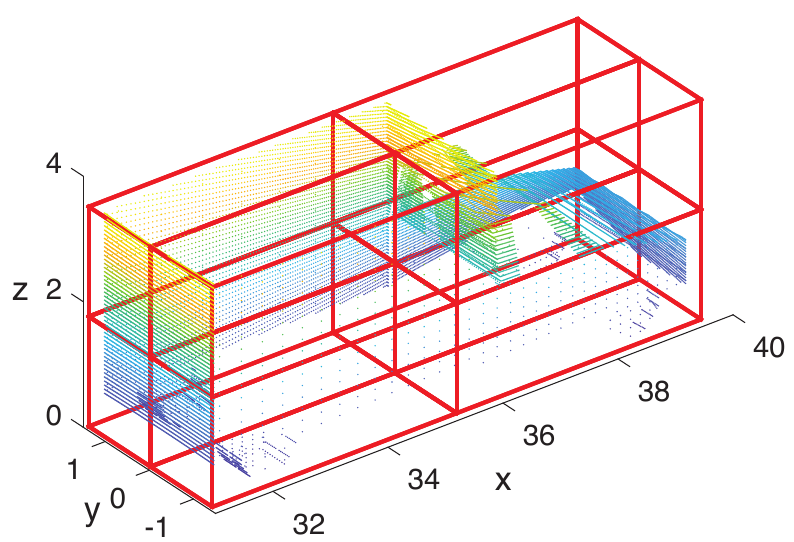}
        \caption{An illustration of 3D bounding box partition with $K=2$.}
\label{fig:cp}
\end{figure}

\emph{Data Quality Indicator}: To coarsely characterize the spatial distribution of object sensing data $\mathcal{D}_n^{(m)}$ of CAV $n$ for object $m$, we partition the 3D cuboid bounding box for object $m$ into $K^3$ disjoint sub-regions, where $K$ denotes the partition resolution along each axis.
Fig. \ref{fig:cp} illustrates the bounding box partition for a truck with $K=2$. 
The lengths of each sub-region along the $x$, $y$, and $z$ axes for object $m$ are $\frac{l_m^{(x)}}{K}$, $\frac{l_m^{(y)}}{K}$ and $\frac{l_m^{(z)}}{K}$, respectively.
Based on the bounding box partition, we define a data quality indicator, $\boldsymbol{Z}_n^{(m)}\in\mathbb{R}^{K^3}$, to characterize the number and spatial distribution of observation points in object sensing data $\mathcal{D}_n^{(m)}$. 
The $\kappa$-th ($1\leq \kappa\leq K^3$) element, $Z_{n,\kappa}^{(m)}$, in vector $\boldsymbol{Z}_n^{(m)}$ denotes the number of observation points from  $\mathcal{D}_n^{(m)}$ that are located inside the $\kappa$-th sub-region. 
Typically, if data quality indicator $\boldsymbol{Z}_n^{(m)}$ contains more non-zero elements and has a larger value for each non-zero element, there are more observation points spreading over the surface of object $m$, leading to a potentially higher object classification accuracy.

\subsection{Task Model}\label{sec:Task Model}

The perception task is executed in two phases: bounding box detection and object classification.

Bounding box detection is to detect the existence of objects in the RoI and have an accurate bounding box estimation for each object.
Typically, low-resolution sensing data can provide sufficient accuracy for bounding box detection~\cite{arnold2020cooperative}. 
To improve the detection accuracy, we fuse the low-resolution sensing data of all CAVs. 
First, each CAV down-samples the raw sensing data to reduce the data resolution. 
Then, each assisting CAV transmits the low-resolution sensing data to the ego CAV. 
After the ego CAV receives all the low-resolution sensing data, 
it fuses all the received data with its own low-resolution sensing data, and executes a classic lightweight bounding box detection algorithm 
to estimate the 6-tuple bounding box parameters~\cite{qiu2021autocast}.
Finally, the ego CAV transmits the estimated bounding box parameters for all objects to all the assisting CAVs\footnote{The bounding boxes of the CAVs are also detected but can be excluded, as each CAV can accurately localize itself and notifies the ego CAV of its location~\cite{qiu2021autocast}. }.
Considering the low data resolution and lightweight data processing, we ignore the resource cost in bounding box detection. 
Assume that all the $M$ objects are successfully detected without loss of generality. 
With the estimated bounding box parameters, the object sensing data for each object can be extracted from the full raw sensing data at each CAV, and the data quality indicators can be calculated.

Object classification is usually performed using an AI model, such as a convolution neural network (CNN), to estimate the probabilities of each class for an object. 
Consider that a pre-trained object classification AI model is stored at each computing node. 
As the AI model processing requires high-resolution sensing data and consumes considerable computing resources, we focus on the object classification phase for the cooperative sensing and computing scheme. 
The perception task in object classification phase can be partitioned into $M$ parallel subtasks, among which subtask $m$ is to classify object $m$ by using the AI model. Let $a_m$ denote the object classification accuracy for subtask $m$. 
All the subtasks should be completed within delay bound $T$, with an accuracy requirement of $a_m\geq A$ for subtask $m$. 
Each subtask is placed at either the RSU or one of the CAVs. 
Let $\boldsymbol{e} = \{e_n^{(m)}, \forall n \in\mathcal{N}^+, \forall m\in\mathcal{M}\}$ denote a binary subtask placement decision matrix in $\mathbb{R}^{(N+1) \times M}$, with decision variable $e_n^{(m)}=1$ indicating that subtask $m$ is placed at computing node $n$, and $e_n^{(m)}=0$ otherwise. 
The following constraint ensures that subtask $m$ is placed at a single computing node, given by
\begin{align} 
    \sum_{n\in\mathcal{N}^+} e_n^{(m)} = 1, \quad \forall m\in\mathcal{M}.
    \label{eq-placement-single}
\end{align}

For subtask $m$, there are at most $N$ sets of object sensing data provided by $N$ CAVs. 
Let $\boldsymbol{s}  = \{s_n^{(m)}, \forall n \in\mathcal{N}, \forall m\in\mathcal{M}\}$ denote a binary data selection decision matrix in $\mathbb{R}^{N \times M}$, with $s_n^{(m)}=1$ indicating that the object sensing data of CAV $n$ for object $m$, i.e., $\mathcal{D}_n^{(m)}$, are selected for subtask $m$, and $s_n^{(m)}=0$ otherwise. 
Let $\mathcal{N}^{(m)}=\left\{n\in\mathcal{N}\big|s_n^{(m)}=1 \right\}$ denote the set of CAVs whose object sensing data are selected for subtask $m$. 
Then, the total sensing data for subtask $m$, denoted by $\mathcal{D}^{(m)}$, are a fusion of all the selected object sensing data from CAVs in set $\mathcal{N}^{(m)}$, given by 
\begin{align} 
    \mathcal{D}^{(m)} = \cup_{n\in\mathcal{N}^{(m)}} \mathcal{D}_n^{(m)}.   
    \label{eq-data-fusion}
\end{align}
As $\mathcal{D}_n^{(m)}$ is a set of 3D coordinates for CAV $n$'s observation points located inside the bounding box of object $m$, the data fusion corresponds to a union operation in~\eqref{eq-data-fusion}. 

Let $\boldsymbol{Z}^{(m)}$ be a data quality indicator of fused object sensing data $\mathcal{D}^{(m)}$ for subtask $m$, given by
\begin{align} 
    \boldsymbol{Z}^{(m)} = \sum_{n\in\mathcal{N}} s_n^{(m)} \boldsymbol{Z}_n^{(m)}
    \label{eq-fused-data-quality-vector}
\end{align}
where $\boldsymbol{Z}_n^{(m)}$ is the data quality indicator for object sensing data $\mathcal{D}_n^{(m)}$. 
If subtask $m$ is placed at computing node $n$, i.e., $e_n^{(m)}=1$, all the selected object sensing data for subtask $m$ should either be transmitted to or be locally available at computing node $n$. In the latter case, computing node $n$ is a CAV.   
When all the selected object sensing data are available at the computing node, the data are fused and then processed by using the object classification AI model. 

\subsection{Computing Model}

For 3D point cloud processing by the object classification AI model, the total computing demand increases roughly proportionally with the data volume~\cite{zhang2021emp}. Let $\epsilon$ denote the computation intensity (in cycle/point) representing the average number of CPU cycles for computing one observation point in the sensing data. 
Then, the computing demand (in CPU cycles) of subtask $m$ for processing the fused object sensing data, denoted by $\mu^{(m)}$, is given by
\begin{align} 
\mu^{(m)} = \epsilon \sum_{n\in\mathcal{N}} s_n^{(m)} \big|\mathcal{D}_n^{(m)}\big|, \quad \forall m\in\mathcal{M}
\label{eq-layer-comp-demand}
\end{align}
where $\big|\mathcal{D}_n^{(m)}\big|$ is the number of observation points in object sensing data $\mathcal{D}_n^{(m)}$.
Note that the computing demand of the data fusion operation in~\eqref{eq-data-fusion} is negligible.   

Each computing node can support multiple subtasks.
The total computing demand for all the subtasks placed at computing node $n\in\mathcal{N}^+$, denoted by $\mu_n$, is given by
\begin{align} 
        \mu_n = \sum_{m\in\mathcal{M}} e_n^{(m)} \mu^{(m)}, \quad \forall n\in\mathcal{N}^+.
    \label{eq-comp-demand}
\end{align}
Let $f_n$ denote the amount of available computing resources (in cycle/s or Hz) at computing node $n$. 
Let $\boldsymbol{\alpha} =\{\alpha_n, \forall n\in\mathcal{N}^+ \}$ be a continuous decision vector in $\mathbb{R}^{N+1}$, where $\alpha_n$ is the fraction of computing resource usage at computing node $n$. 
For each subtask placed at computing node $n$, the data processing can start only after the selected sets of object sensing data from different CAVs are available at the computing node. 
For simplicity, assume that the AI model processing at a computing node starts once the computing node receives all data for all the assigned subtasks.
Then, the total computing time for all the subtasks placed at computing node $n$, denoted by $t_n$, is calculated as
\be 
t_n =  \left\{ 
\begin{array}{rcl}
{\frac{\mu_n}{ \alpha_n f_n}   \text{,} \hspace{0cm}}   & { \text{ if } \alpha_n>0\hspace{0.1cm}} \\
{0   \text{,} \hspace{0.48cm}}  &  { \text{ if } \alpha_n=0.\hspace{0cm}} 
\label{eq-comp-delay-n}
\end{array} \right.
\ee

\subsection{Communication Model}

If CAV $n$ provides object sensing data $\mathcal{D}_n^{(m)}$ for subtask $m$, i.e, $s_n^{(m)}=1$, but subtask $m$ is placed at computing node $n'\in\mathcal{N}^+\backslash\{n\}$, i.e., $e_{n'}^{(m)}=1$, object sensing data $\mathcal{D}_n^{(m)}$ should be transmitted from CAV $n$ to computing node $n'$.
Let $\varphi$ denote the data size (in bit) of one observation point.  
Then, the total size of the sensing data transmitted from CAV $n$ to computing node $n'$ (potentially for different subtasks), denoted by $\rho_{n,n'}$, 
is given by
\begin{align} 
      \hspace{-2mm}  \rho_{n,n'} = \varphi \sum_{m\in\mathcal{M}} s_n^{(m)} e_{n'}^{(m)}  \big|\mathcal{D}_n^{(m)}\big|, 
    \forall n\in\mathcal{N}, 
    \forall n' \hspace{-1mm}\in\hspace{-1mm}\mathcal{N}^+\backslash\{n\}.
    \label{eq-data-size}
\end{align}

Orthogonal frequency division multiplexing (OFDM) is employed for V2X transmissions. 
Consider that all the V2I and V2V links share a spectrum with total bandwidth $B$.
Let $\boldsymbol{\beta} =\{\beta_{n,n'}, \forall n\in\mathcal{N}, \forall n'\in\mathcal{N}^+ \}$ denote a continuous bandwidth allocation decision matrix in $\mathbb{R}^{N \times (N+1)}$, where $\beta_{n,n'}$ represents the fraction of bandwidth allocated to the communication link from CAV $n$ to computing node $n'$.  
We have $\beta_{n,n}\equiv0$ for $n\in\mathcal{N}$.
The total allocated fraction of bandwidth should not exceed 1, given by
\begin{align} 
        \sum_{n\in\mathcal{N}}\sum_{n'\in\mathcal{N}^+} \beta_{n,n'} \leq 1.
    \label{eq-comm-capacity}
\end{align}
The average transmission rate between CAV $n$ and computing node $n'$ is given by
\begin{align} 
        R_{n,n'} = \beta_{n,n'}B\log_2 \left(1+\frac{P_n \big|h_{n,n'}\big|^2d_{n,n'}^{-\gamma}}{\sigma^2}\right)
    \label{eq-delta-r-mid}
\end{align}
where $P_n$ denotes the transmit power of CAV $n$, $h_{n,n'}$ is the channel fading coefficient from CAV $n$ to computing node $n'$, $d_{n,n'}$ is the distance between CAV $n$ and computing node $n'$, $\gamma$ is the path loss exponent, and $\sigma^2$ represents the received noise power.
Here, we assume constant distance $d_{n,n'}$ during the perception task period, with the consideration of low latency requirement for the perception task, e.g., $T = 20ms$. Then, the average transmission time for all the sensing data transmitted from CAV $n$ to computing node $n'$, denoted by $t_{n,n'}$, is given by
\be 
t_{n,n'} =  \left\{ 
\begin{array}{rcl}
{\frac{\rho_{n,n'}}{ R_{n,n'} }   \text{,} \hspace{0cm}}   &  { \text{ if } \beta_{n,n'}>0\hspace{0.06cm}} \\
{0   \text{,} \hspace{0.58cm}}  &  { \text{ if } \beta_{n,n'}=0.\hspace{0cm}}
\label{eq-trans-delay-nn}
\end{array} \right.
\ee

\subsection{Generalization}

We focus on one perception task of an ego CAV and ignore the perception tasks of assisting CAVs. 
The target scenario of the proposed cooperative sensing and computing scheme is not limited to the considered simplified case. When multiple nearby CAVs with overlapping RoIs have perception tasks, we can consider an augmented perception task in a union RoI of all CAVs. 
The proposed scheme can be extended to support such an augmented perception task. 
Each object in the union RoI is classified once at an assigned computing node. Then, the object classification results can be shared among all CAVs according to their interests in the individual RoIs, with minimal transmission cost.

\section{Joint Data Selection, Subtask Placement and Resource Allocation Problem} 
\label{sec:Joint Data Selection, Subtask Placement and Resource Allocation Problem}

\subsection{Problem Definition}

To support the perception task initiated by the ego CAV, the assisting CAVs can cooperate with the ego CAV for both sensing and computing. If assisting CAV $n$ provides object sensing data for at least one subtask, i.e., $\sum_{m\in\mathcal{M}} s_n^{(m)}\geq 1$, it participates in the \emph{cooperative sensing} with the ego CAV; if at least one subtask is placed at assisting CAV $n$, i.e., $\sum_{m\in\mathcal{M}} e_n^{(m)}\geq 1$, it participates in the \emph{cooperative computing}.
The RSU can also be involved in the cooperative computing by supporting the computation of at least one subtask. 

For the accuracy satisfaction of an object classification subtask, it is preferable to select object sensing data from more CAVs especially from those with complementary viewing angles, to enhance the data quality of the fused object sensing data for a higher data intensity and a more even spatial distribution. 
As a better data quality leads to a higher object classification accuracy, data selection from more CAVs potentially improves the accuracy at the cost of more communication resources for transmitting the selected object sensing data and more computing resources for processing the fused object sensing data.
However, adding data from more CAVs suffers from a diminishing marginal accuracy gain with almost linearly increasing resources.   
For network resource efficiency, 
it is necessary to select the best group of CAVs for each subtask to provide the minimum amount of object sensing data that satisfy the accuracy requirement.  
For the subtask placement, distributed computing among the RSU and CAVs not only relieves the computation load at the edge server, enhancing the delay performance, but also potentially reduces the communication cost by placing a subtask and selecting the corresponding object sensing data at the same CAV.  
Hence, for delay satisfaction and resource efficiency, the data selection and subtask placement decisions should be jointly determined. 

To support the cooperative sensing and computing scheme with resource efficiency and quality of service (QoS) satisfaction in both accuracy and delay, 
we study a joint sensing data selection, subtask placement, and resource allocation problem for edge-assisted CAVs.  
We want to determine the sensing data selection and placement for multiple parallel object classification subtasks, with efficient V2I and V2V transmission resource allocation among multiple CAVs and RSU, and with a minimum total computing resource usage at all computing nodes.
Decision variables include binary data selection decision matrix, $\boldsymbol{s}\in\mathbb{R}^{N \times M}$, binary subtask placement decision matrix, $\boldsymbol{e}\in \mathbb{R}^{(N+1) \times M}$, continuous computing resource usage decision vector, $\boldsymbol{\alpha} \in\mathbb{R}^{N+1}$, and continuous bandwidth allocation decision matrix, $\boldsymbol{\beta} \in\mathbb{R}^{N \times (N+1)}$. 
As the object classification accuracy for each subtask depends on the data selection decision, we profile an accuracy estimation function to facilitate accuracy-aware data selection by using a supervised learning model as presented in Subsection~\ref{sec:DNN-Based Classification Accuracy Prediction}, based on which the joint data selection, subtask placement and resource allocation problem is formulated in Subsection~\ref{sec:formulation}.

\subsection{Learning-Based Accuracy Estimation}
\label{sec:DNN-Based Classification Accuracy Prediction}

For $L$-class object classification, the object classification result is an $L$-dimension estimated class probability vector.
A higher estimated probability for the true class indicates a more confident estimation by the AI model and implies a higher accuracy~\cite{10137743}. 
For simplicity, we use the estimated true class probability as the classification accuracy metric. 
For object $m$, the ground-truth classification accuracy, $a_m$, fully depends on fused object sensing data $\mathcal{D}^{(m)}$ and the object classification AI model, which is unknown until fused object sensing data $\mathcal{D}^{(m)}$ are processed by the object classification AI model. 
Hence, for data selection decision with ensured accuracy before data processing, accuracy estimation is required for the AI-based object classification based on any fused object sensing data.

Considering an object classification AI model with pre-trained parameters, the dominant factor that impacts the classification accuracy of object $m$ is the quality of fused object sensing data $\mathcal{D}^{(m)}$, which is characterized by data quality indicator $\boldsymbol{Z}^{(m)}$. 
As an object with smaller size tends to require less observation points for accurate classification, the bounding box dimensions of the objects, i.e., $\{l_m^{(x)},l_m^{(y)},l_m^{(z)}\}$ for object $m$, should also be considered in the accuracy estimation. 
Accordingly, we profile an accuracy estimation function by a supervised learning model, specifically a deep neural network (DNN) model, with data quality indicator and bounding box dimensions as inputs and estimated object classification accuracy as output, represented as
\begin{align} 
    \hat{a}_m = \mathsf{f}^{DNN}\left(\boldsymbol{Z}^{(m)}, l_m^{(x)},l_m^{(y)},l_m^{(z)}\right), \quad \forall m\in\mathcal{M}
    \label{eq-accuracy-estimation}
\end{align}
where $\hat{a}_m$ denotes the estimated classification accuracy for object $m$. 
The accuracy estimation learning model has an input dimension of $(K^3 + 3)$ and an output dimension of one,
which can be pre-trained offline with simulated point cloud data and refined online with real-world data. 

To offline pre-train the accuracy estimation learning model, we create a training dataset using simulated point cloud data in random autonomous driving scenarios with a random number of CAVs equipped with LiDAR sensors and a random number of objects distributed on a road. 
For each simulated scenario, raw LiDAR point clouds are generated at the CAVs, which are randomly fused to generate more simulated LiDAR point clouds with different point number and spatial distribution. 
From the point clouds, we \emph{extract} per-object sensing data instances using bounding box detection algorithms, \emph{calculate} the data quality indicators for all data instances based on  bounding box partition, \emph{process} the data instances using an object classification AI model, and \emph{obtain} the ground-truth object classification accuracy for each data instance.
Each data instance corresponds to one training data sample, which is a $(K^3 + 3)$-dimension vector concatenated from a $K^3$-dimension data quality indicator and three bounding box dimension values (including length, width, height). The label for each training data sample is the corresponding ground-truth object classification accuracy. 
With the training dataset, we train the accuracy estimation learning model by minimizing the mean squared error (MSE) between the estimated and ground-truth object classification accuracy for all the training data samples.

\subsection{Problem Formulation}
\label{sec:formulation}

Based on the accuracy estimation learning model, the object classification accuracy for subtask $m$ can be estimated for any candidate data selection decision.
For accuracy satisfaction, the estimated accuracy of each subtask should satisfy accuracy requirement $A$, given by
\begin{align} 
        \mathsf{f}^{DNN}\left(\sum_{n\in\mathcal{N}} s_n^{(m)} \boldsymbol{Z}_n^{(m)}, l_m^{(x)},l_m^{(y)},l_m^{(z)}\right) \geq A, \forall m\in\mathcal{M}
    \label{eq-accuracy-constr}
\end{align}
according to \eqref{eq-fused-data-quality-vector} and \eqref{eq-accuracy-estimation}.
In \eqref{eq-accuracy-constr}, $\boldsymbol{Z}_n^{(m)}$, $l_m^{(x)}$, $l_m^{(y)}$, $l_m^{(z)}$ are known parameters during the object classification phase.

Assume that all CAVs selected to send at least one instance of object sensing data to other computing nodes start their data transmissions simultaneously.
Then, any subtask placed at computing node $n'$ can be completed within at most a time of $\left(\max_{n'\in\mathcal{N}} t_{n,n'} \right) + t_{n'}$.
As the completion time for any subtask at any computing node in $\mathcal{N}^+$ should not exceed delay bound $T$, we have a delay constraint as 
\begin{align} 
        \max_{n'\in\mathcal{N}^+}{  \left( \left(\max_{n\in\mathcal{N}} t_{n,n'} \right) + t_{n'} \right)}  \leq T
    \label{eq-enhanced-comp-delay}
\end{align} 
which is equivalent to
\begin{align} 
        t_{n,n'}  + t_{n'} \leq T, \quad \forall n\in\mathcal{N}, ~\forall n'\in\mathcal{N}^+.
    \label{eq-delay-constr}
\end{align}

Let $\boldsymbol{\chi} =\{\chi_{n,n'}, \forall n\in\mathcal{N}, \forall n'\in\mathcal{N}^+ \}$ be an auxiliary binary decision matrix in $\mathbb{R}^{N \times (N+1)}$, with $\chi_{n,n'}=1$ indicating that the wireless communication (either V2I or V2V) link between CAV $n$ and computing node $n'$ is activated for data transmission when CAV $n$ sends its object sensing data for at least one subtask to computing node $n'$, and $\chi_{n,n'}=0$ otherwise. 
We have $\chi_{n,n}\equiv0$ for $n\in\mathcal{N}$.
The relationship among $\boldsymbol{\chi}$, data selection decision $\boldsymbol{s}$, and subtask placement decision $\boldsymbol{e}$ is given by
\begin{align} 
        \frac{ \sum_{m\in\mathcal{M}}{s_n^{(m)}e_{n'}^{(m)}} }{M} \leq \chi_{n,n'} \leq \sum_{m\in\mathcal{M}}{s_n^{(m)}e_{n'}^{(m)}}, \nonumber \\ 
        \forall n\in\mathcal{N}, ~\forall n'\in\mathcal{N}^+.
    \label{eq-link-activation}
\end{align}
We have $\chi_{n,n'}=0$ for $\sum_{m\in\mathcal{M}}{s_n^{(m)}e_{n'}^{(m)}}=0$, 
and $\chi_{n,n'}=1$ otherwise.
Under the assumption that each CAV is equipped with one half-duplex transceiver radio, at most one communication link starting or ending at CAV $n\in\mathcal{N}$ can be activated for data transmission, given by
\begin{align} 
        \sum_{n'\in\mathcal{N}^+\backslash \{n\}}{\chi_{n,n'}} + \sum_{n'\in\mathcal{N}\backslash \{n\}}{\chi_{n',n}}\leq 1, \quad 
        \forall n\in\mathcal{N}.
    \label{eq-half-duplex}
\end{align}
Note that \eqref{eq-half-duplex} does not hold for computing node $n=N$ (i.e., the RSU), as multiple V2I links can be activated simultaneously for concurrent V2I data transmissions 
due to multiple transceiver radios at the RSU.

To minimize the total resource consumption cost for supporting the cooperative sensing and computing scheme, which is a weighted summation of the total fractions of transmission and computation resource usage\footnote{We focus on
the fractions rather than the absolute values of total resources for both transmission and computation. },  
\color{black}we formulate a joint data selection, subtask placement, and resource allocation problem as an optimization problem, given by
\begin{align*}
        \mathbf{P}1: ~\min_{\boldsymbol{e}, \boldsymbol{s}, \boldsymbol{\beta}, \boldsymbol{\alpha}, \boldsymbol{\chi}}
        & \ ~ \omega \sum_{n\in\mathcal{N}}\sum_{n'\in\mathcal{N}^+}\beta_{n,n'} + \left(1-\omega\right) \frac{\sum_{n\in\mathcal{N}^+} \alpha_n f_n}{\sum_{n\in\mathcal{N}^+} f_n} \\
        \textrm{s.t. } \hspace{0.2cm}
        & \quad \eqref{eq-placement-single}, \eqref{eq-comm-capacity}, \eqref{eq-accuracy-constr}, \eqref{eq-delay-constr}, \eqref{eq-link-activation}, \eqref{eq-half-duplex} \\
        & \quad s_n^{(m)} \in \{0, 1\}, \quad \forall m\in\mathcal{M}, ~\forall n\in\mathcal{N} \\
        & \quad e_n^{(m)} \in \{0, 1\}, \quad \forall m\in\mathcal{M}, ~\forall n\in\mathcal{N}^+ \\
        & \quad \chi_{n,n'} \in \{0, 1\}, \quad \forall n\in\mathcal{N}, ~\forall n'\in\mathcal{N}^+ \\
        & \quad \chi_{n,n} = \beta_{n,n} =0, \quad \forall n\in\mathcal{N} \\
        & \quad 0 \leq \beta_{n,n'} \leq 1, \quad \forall n\in\mathcal{N}, ~\forall n'\in\mathcal{N}^+ \\
        & \quad 0 \leq \alpha_n \leq 1, \quad \forall n\in\mathcal{N}^+.
\end{align*}
Problem $\mathbf{P}1$ has constraints in terms of topology, accuracy, delay, and resource capacity.
The topology constraints include the single computing node placement constraint for each subtask in \eqref{eq-placement-single} and the half-duplex communication constraints of each CAV in \eqref{eq-link-activation} and \eqref{eq-half-duplex}. 
The accuracy and delay constraints are given by \eqref{eq-accuracy-constr} and \eqref{eq-delay-constr}, respectively.
Constraint \eqref{eq-comm-capacity} corresponds to the transmission resource capacity. 
The remaining constraints are range requirements for the decision variables. 
Weight $\omega\in(0,1)$ balances between the transmission and computing resource consumption.

\subsection{Practical Implementation}

In the considered edge-assisted autonomous driving scenarios, network devices including both CAVs and RSU have communication capabilities and serve as computing nodes, while CAVs also serve as sensing devices. 
When the ego CAV moves, the nearby assisting CAVs and the communicating RSU would change over time, and it is possible that the ego CAV moves into an area without RSU coverage or assisting CAVs. 
For seamless and reliable control, a controller, that coordinates the cooperative sensing and computing among all the network devices, can be placed at the ego CAV that initializes the perception task. 

The controller is responsible for three control subtasks: 1) bounding box detection, which requires the lightweight low-resolution sensing data from CAVs, 2) learning-based accuracy estimation, which requires the data quality indicators calculated by CAVs and the bounding box dimensions, and 3) joint data selection, subtask placement, and resource allocation decision, which requires the estimated accuracy and some extra context information, e.g., network topology, channel conditions, resource availability and QoS requirements.  
The bounding box detection requires the point clouds of different CAVs be aligned with a global coordinate system, which is facilitated by CAV positioning based on global positioning system (GPS) or simultaneous localization and mapping (SLAM) techniques~\cite{hu2022where2comm,chiang2020performance}. 
The accuracy estimation learning model can be online updated based on continual learning techniques, to accommodate the environment evolution over time such as in terms of object types~\cite{jia2021cost}. 
All the control information can be collected via a dedicated control channel. 
As the data size of the control information is small in comparison with that of the object sensing data, the signalling overhead is negligible.

\section{Problem Solution}
\label{sec:Problem Solution}

As the accuracy constraint in \eqref{eq-accuracy-constr} incorporating DNN-based accuracy estimation does not have a closed-form expression, Problem $\mathbf{P}1$ is intractable. 
We notice that the accuracy and topology constraints depend only on the binary decision variables including data selection and subtask placement, while the delay and resource capacity constraints depend on the joint decisions of data selection, subtask placement, and resource allocation.
Accordingly, we propose an iterative solution to problem $\mathbf{P}1$. 
In the solution, an outer module iteratively optimizes the data selection and subtask placement based on a genetic algorithm, which relies on an inner module to check the feasibility and cost. In each iteration, the outer module provides the inner module with a candidate data selection and subtask placement solution, denoted by $(\boldsymbol{s},\boldsymbol{e})$, which is feasible in terms of the accuracy and topology constraints.
Given an $(\boldsymbol{s},\boldsymbol{e})$ pair, the inner module optimizes the resource allocation for a minimal total  resource consumption cost that satisfies the delay and resource capacity constraints, by solving a resource allocation subproblem.
If the subproblem is infeasible, 
$(\boldsymbol{s},\boldsymbol{e})$ is infeasible in terms of delay and resource constraints. 
Otherwise, $(\boldsymbol{s},\boldsymbol{e})$ is feasible in terms of all constraints in problem $\mathbf{P}1$, and a cost provided by the inner module corresponding to $(\boldsymbol{s},\boldsymbol{e})$ should be evaluated by the outer module.

\subsection{Resource Allocation Subproblem}
\label{sec:Resource Allocation Subproblem}

Given an $(\boldsymbol{s},\boldsymbol{e})$ pair, the auxiliary binary link activation decision matrix, $\boldsymbol{\chi}\in\mathbb{R}^{N \times (N+1)}$, is determined according to \eqref{eq-link-activation}.
Given link activation status $\boldsymbol{\chi}$,  
we consider a network topology represented as a directed graph $G=\{\mathcal{N}^{\mathtt{A}},\mathcal{L}^{\mathtt{A}}\}$, 
where $\mathcal{N}^{\mathtt{A}} \subset \mathcal{N}^+$ is a set of nodes composed of the starting and ending nodes (either CAV or RSU) of all the activated links, 
and $\mathcal{L}^{\mathtt{A}}$ is a set of directed links composed of all the activated links.
For nodes $n\in\mathcal{N}$ and $n'\in\mathcal{N}^+$, let $(n,n')$ denote a directed link from node $n$ to node $n'$. We have $n\in\mathcal{N}^{\mathtt{A}}$, $n'\in\mathcal{N}^{\mathtt{A}}$ and $(n,n')\in\mathcal{L}^{\mathtt{A}}$ if link $(n,n')$ is activated, i.e., $\chi_{n,n'}=1$.
Let $\boldsymbol{\alpha}^{\mathtt{A}} =\{\alpha_n, \forall n\in\mathcal{N}^{\mathtt{A}} \}$ denote the fractions of computing resource usage at the nodes in set $\mathcal{N}^{\mathtt{A}}$, 
and let $\boldsymbol{\beta}^{\mathtt{A}} =\{\beta_{n,n'}, \forall (n,n')\in\mathcal{L}^{\mathtt{A}}\}$ denote the fractions of bandwidth allocated to the links in set $\mathcal{L}^{\mathtt{A}}$.  
Let constant $C_n = \frac{\mu_n}{f_n}$ denote the computing time at node $n$ using all the available computing resources, i.e., for $\alpha_n=1$.  
Let constant $C_{n,n'} = \frac{\rho_{n,n'}}{B\log_2 \left(1+ P_n |h_{n,n'}|^2 d_{n,n'}^{-\gamma} \big/ \sigma^2\right)}$ denote the transmission time over link $(n,n')$ using whole bandwidth $B$, i.e., for $\beta_{n,n'}=1$. 
The resource allocation subproblem for the inner module is to minimize the total resource consumption cost by determining the resource allocation decision variables, $\boldsymbol{\alpha}^{\mathtt{A}}$ and $\boldsymbol{\beta}^{\mathtt{A}}$, while satisfying delay and resource capacity constraints, given by 
\begin{align}
        \mathbf{P}2: \quad \min_{\boldsymbol{\alpha}^{\mathtt{A}}, \boldsymbol{\beta}^{\mathtt{A}}}
        & \ \quad \omega \sum_{(n,n')\in\mathcal{L}^{\mathtt{A}}} \beta_{n,n'} + \left(1-\omega\right) \frac{\sum_{n\in\mathcal{N}^{\mathtt{A}}} \alpha_n f_n}{\sum_{n\in\mathcal{N}^+} f_n} \nonumber\\
        \textrm{s.t. } \hspace{0cm}
        & \quad \sum_{(n,n')\in\mathcal{L}^{\mathtt{A}}} \beta_{n,n'} \leq 1 \nonumber\\
        & \quad \frac{C_{n,n'}}{\beta_{n,n'}} + \frac{C_{n'}}{\alpha_{n'}} \leq T, \quad \forall (n,n')\in\mathcal{L}^{\mathtt{A}} 
        \label{eq-delay-original}
        \\
        & \quad 0 < \beta_{n,n'} \leq 1, \quad \forall (n,n')\in\mathcal{L}^{\mathtt{A}} \nonumber\\
        & \quad 0 < \alpha_n \leq 1, \quad \forall n\in\mathcal{N}^{\mathtt{A}} \nonumber
\end{align}
where \eqref{eq-delay-original} is rewritten based on \eqref{eq-comp-delay-n}, \eqref{eq-trans-delay-nn} and \eqref{eq-delay-constr}.
Problem $\mathbf{P}2$ is a convex optimization problem with a linear objective function and convex inequality constraints. 
However, as \eqref{eq-delay-original} involves the division by decision variables, problem $\mathbf{P}2$ cannot be directly solved by an optimization solver such as Gurobi~\cite{gurobi,10137743}. 
We can transform the problem to a second-order cone programming (SOCP) problem with zero optimality gap, by transforming \eqref{eq-delay-original} to two rotated second-order cone constraints of dimension $2$ for $\forall (n,n')\in\mathcal{L}^{\mathtt{A}}$, which can be solved in polynomial time~\cite{gurobi,10137743,lobo1998applications}. 

Let $\tau\left(\boldsymbol{s},\boldsymbol{e}\right)$ be a binary feasibility flag for problem $\mathbf{P}2$ given $(\boldsymbol{s},\boldsymbol{e})$, with $\tau\left(\boldsymbol{s},\boldsymbol{e}\right)=1$ if the problem is feasible and $\tau\left(\boldsymbol{s},\boldsymbol{e}\right)=0$ otherwise. Let $o^{\ast}\left(\boldsymbol{s},\boldsymbol{e}\right)$ be the optimal objective value of problem $\mathbf{P}2$ given $(\boldsymbol{s},\boldsymbol{e})$, corresponding to the minimal total resource consumption cost with optimal resource allocation.
Note that $o^{\ast}\left(\boldsymbol{s},\boldsymbol{e}\right)$ is defined only when problem $\mathbf{P}2$ is feasible given $(\boldsymbol{s},\boldsymbol{e})$.

\subsection{Genetic Algorithm}
\label{sec:Genetic Algorithm}

The outer module jointly optimizes the data selection decision, $\boldsymbol{s}$, and the subtask placement decision, $\boldsymbol{e}$, by solving an optimization problem given by
\begin{align*}
        \mathbf{P}3: \quad \min_{\boldsymbol{e}, \boldsymbol{s}, \boldsymbol{\chi}}
        & \ \quad o^{\ast}\left(\boldsymbol{s},\boldsymbol{e}\right) \nonumber\\
        \textrm{s.t. } \hspace{0cm}
        & \quad \eqref{eq-placement-single}, \eqref{eq-accuracy-constr}, \eqref{eq-link-activation}, \eqref{eq-half-duplex} \nonumber\\
        & \quad \tau\left(\boldsymbol{s},\boldsymbol{e}\right)=1 
        \\
        & \quad s_n^{(m)} \in \{0, 1\}, \quad \forall m\in\mathcal{M}, ~\forall n\in\mathcal{N} \nonumber\\
        & \quad e_n^{(m)} \in \{0, 1\}, \quad \forall m\in\mathcal{M}, ~\forall n\in\mathcal{N}^+ \nonumber\\
        & \quad \chi_{n,n'} \in \{0, 1\}, \quad \forall n\in\mathcal{N}, ~\forall n'\in\mathcal{N}^+ \nonumber\\
        & \quad \chi_{n,n} =0, \quad \forall n\in\mathcal{N} \nonumber
\end{align*}
where $o^{\ast}\left(\boldsymbol{s},\boldsymbol{e}\right)$ and $\tau\left(\boldsymbol{s},\boldsymbol{e}\right)$ are obtained by solving resource allocation subproblem $\mathbf{P}2$ given $(\boldsymbol{s},\boldsymbol{e})$ in the inner module.

To solve problem $\mathbf{P}3$, we propose a genetic algorithm (GA) which gradually optimizes the joint data selection and subtask placement decision by iteratively selecting candidate solutions (referred to as individuals in GA) with lower costs from each generation to reproduce new individuals in a next generation~\cite{sun2018cooperative,9416871}. 
Let integer $k$ be the generation index. 
Each generation is composited by $J$ individuals. 
Let $\mathcal{J}=\{0,\cdots,J-1\}$ denote the individual index set for each generation. 
Let $\mathbf{V}^{k,j}$ denote the $j$-th individual in generation $k$, corresponding to the $j$-th candidate joint data selection and subtask placement solution, $\left(\boldsymbol{s}^{k,j},\boldsymbol{e}^{k,j}\right)$.
The population in generation $k$ is represented as $\boldsymbol{\Phi}^{k}=\left\{\mathbf{V}^{k,j}, ~\forall j\in\mathcal{J}\right\}$.
The cost of individual $\mathbf{V}^{k,j}$, denoted by $o^{k,j}$, is given by $o^{k,j} = o^{\ast}\left(\boldsymbol{s}^{k,j},\boldsymbol{e}^{k,j}\right)$, which is the minimal total resource consumption cost obtained by solving resource allocation subproblem $\mathbf{P}2$ given $\left(\boldsymbol{s}^{k,j},\boldsymbol{e}^{k,j}\right)$.

In GA, an individual is usually represented by a sequence of genes.
We consider $M$ genes for each individual, i.e., $\mathbf{V}^{k,j} = \left\{ \boldsymbol{v}_m^{k,j}, ~\forall m\in \mathcal{M}\right\}$, where 
the $m$-th gene, $\boldsymbol{v}_m^{k,j}$, is a concatenated data selection and subtask placement decision vector for subtask $m$, denoted by 
\begin{align} 
    \boldsymbol{v}_m=\left[\boldsymbol{s}_m, \boldsymbol{e}_m\right], \quad \forall m\in\mathcal{M}
\end{align}
with $\boldsymbol{s}_m= \{s_n^{(m)}, \forall n \in \mathcal{N}\}$ and $\boldsymbol{e}_m= \{e_n^{(m)}, \forall n \in \mathcal{N}^+\}$.

\begin{algorithm}[t] 
\caption{Genetic algorithm based iterative solution.}
\label{Alg1}
\begin{algorithmic}[1]
\State {\textbf{Initialization}: Generate an initial population containing $J$ random feasible individuals for generation $k=0$;}
\For{each generation with $0\leq k\leq \Gamma-1$}
  \State {The elite individual with the minimal cost in }
  \Statex \quad\quad{generation $k$ survives to generation $k+1$;}
  \For{$ 1 \leq j \leq J-1$}
    \State {\textbf{Selection}: Randomly select two individuals $\mathbf{V}^1$}
    \Statex \quad\quad\quad~{ and $\mathbf{V}^2$ from generation $k$  based on the}
    \Statex \quad\quad\quad~~{cost-dependent selection probabilities;}
    \State {$\hat{\mathbf{V}}=\mathbf{V}^1$;}
    \State {Generate a random number $\xi\in[0,1]$;}
    \If{$\xi \leq p_C$}
      \State {$\hat{\mathbf{V}} = Crossover(\mathbf{V}^1, \mathbf{V}^2)$;}
    \EndIf
    \If{$\xi \leq p_M$}
      \State {$\hat{\mathbf{V}} = Mutation(\hat{\mathbf{V}})$;}
    \EndIf
    \If{Candidate individual $\hat{\mathbf{V}}$ is feasible}
      \State {$\mathbf{V}^{k+1,j}= \hat{\mathbf{V}}$, and $o^{k+1,j}=\hat{o}$;}
    \Else
      \State {$\mathbf{V}^{k+1,j}= \mathbf{V}^1$, and $o^{k+1,j}=o^1$ by default.}
    \EndIf
    \State {$k=k+1$;}
  \EndFor
\EndFor
\end{algorithmic}
\end{algorithm}

\begin{figure}
\centering
\includegraphics[width=1\linewidth]{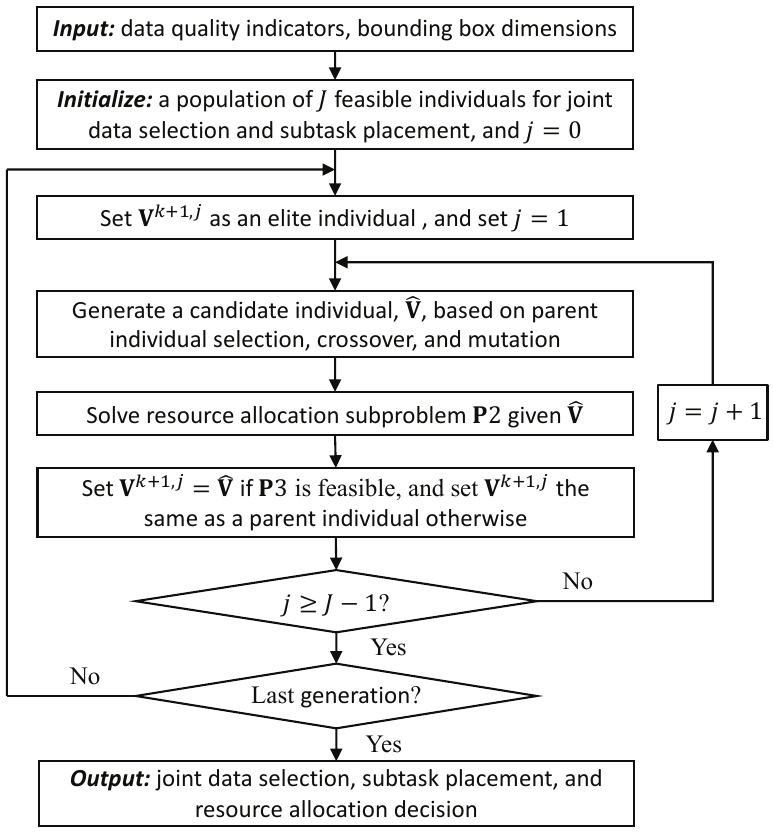}
\caption{A flowchart of the proposed algorithm.}
\label{fig:flow_graph}
\end{figure}

The pseudo codes for the GA algorithm are presented in Algorithm~\ref{Alg1}, and a flowchart of the algorithm is shown in Fig.~\ref{fig:flow_graph}.  
For initialization, population $\boldsymbol{\Phi}^0$ for generation $0$ including $J$ feasible individuals are randomly generated, among which the $j$-th individual, $\mathbf{V}^{0,j}\in\boldsymbol{\Phi}^0$, has cost $o^{0,j}$ (\emph{line} 1). 
Then, the GA algorithm iteratively reproduces a population for each new generation, which includes $J$ individuals with a gradually decreasing average cost, until a maximum number of iterations, $\Gamma$, is reached.

Specifically, at the $k$-th iteration, $J$ new individuals forming population $\boldsymbol{\Phi}^{k+1}$ of generation $k+1$ are reproduced based on population $\boldsymbol{\Phi}^{k}$ of generation $k$.
First, an elite individual $\mathbf{V}^{k,j^*}$ in population $\boldsymbol{\Phi}^{k}$ with the minimum cost, where $j^* = \arg \min_{j\in\mathcal{J}} \hspace{1mm} o^{k,j}$, survives to generation $k+1$ and becomes an individual with index $j=0$ in population $\boldsymbol{\Phi}^{k+1}$, i.e., $\mathbf{V}^{k+1,0}= \mathbf{V}^{k,j^*}$ (\emph{line} 3).
To generate the $j$-th ($j \geq 1$) offspring individual, $\mathbf{V}^{k+1,j}$, in generation $k+1$, two individuals $\mathbf{V}^1 = \left\{ \boldsymbol{v}_m^1,\forall m\in \mathcal{M}\right\}$ and $\mathbf{V}^2 = \left\{ \boldsymbol{v}_m^2,\forall m\in \mathcal{M}\right\}$ in population $\boldsymbol{\Phi}^{k}$ are randomly selected as parents based on cost-dependent selection probabilities (\emph{line} 5). 
Let $p^{k,j}= 1- \frac{ o^{k,j}}{\sum_{j'=0}^{J-1} o_{j'}^{k}}$ be the selection probability for individual $\mathbf{V}^{k,j}$ in generation $k$, 
which indicates that individuals with lower costs have a higher probability to be selected for reproduction.
Given the parent individuals, a candidate individual, $\hat{\mathbf{V}}$, is set to $\mathbf{V}^1$ by default (\emph{line} 6). 
With a probability of $p_C$, a crossover operation is performed between $\mathbf{V}^1$ and $\mathbf{V}^2$, and $\hat{\mathbf{V}}$ is set as the resulting individual (\emph{lines} 8-10).  
With a probability of $p_M$, $\hat{\mathbf{V}}$ is altered with a mutation operation (\emph{lines} 11-13).  
The crossover and mutation operations are described as follows: 
\begin{itemize}

    \item \textbf{Crossover} $-$ Based on a random gene position, $\hat{m}\in \mathcal{M}$, $\hat{\mathbf{V}}$ is generated by inheriting the genes of parent 
    individual $\mathbf{V}^1$ before the random gene position and the genes of parent individual $\mathbf{V}^2$ after the random gene position, given by $\hat{\mathbf{V}} = \left\{ \boldsymbol{v}_m^1, 0\leq  m\ \leq \hat{m}-1\right\} \cup \left\{ \boldsymbol{v}_m^2, \hat{m}\leq  m\ \leq M-1\right\}$; 

    \item  \textbf{Mutation} $-$ Individual $\hat{\mathbf{V}}$ is altered by replacing one gene at a random position by a random gene.

\end{itemize}
Typically, a higher crossover probability, $p_C$, provides more exploration of new areas in a global search space, while a lower mutation probability, $p_M$, helps to avoid excessive local exploration and prevent random disruptive changes that may harm the population's overall quality.  
Next, the feasibility of $\hat{\mathbf{V}}$ is determined by checking all the constraints in problem $\mathbf{P}3$. 
If $\hat{\mathbf{V}}$ is feasible, $\mathbf{V}^{k+1,j}$ is set as $\hat{\mathbf{V}}$, and $o^{k+1,j}$ is set as $\hat{o}$ which is the cost of $\hat{\mathbf{V}}$ (\emph{lines} 14-15).
Otherwise, $\mathbf{V}^{k+1,j}$ is set the same as parent $\mathbf{V}^1$, and $o^{k+1,j}$ is set as $o^1$, the cost of $\mathbf{V}^1$, by default (\emph{line} 17). 
The elite individual in the last generation and the associated resource allocation decisions together composite the algorithm solution. 

\textit{Convergence analysis:}
Empirical studies have shown that genetic algorithms converge for a large class of NP-hard problems~\cite{lorenzo2012optimal}.
Moreover, as the number of candidate data selection and subtask placement decisions is finite, the GA-based outer module is always convergent~\cite{xu2018improved}. 
In each outer iteration, an SOCP problem is solved in the inner module. 
Primal-dual interior-point methods are commonly used for solving SOCPs in an iterative manner, which has been proven to exhibit polynomial-time convergence. 
Specifically, for our transformed SOCP problem with $2|\mathcal{L}^{\mathtt{A}}|$ second-order cone constraints based on problem $\mathbf{P}2$, the number of iterations to decrease the duality gap to a constant fraction of itself is upper bounded by $O(\sqrt{2|\mathcal{L}^{\mathtt{A}}|})$~\cite{lobo1998applications}. 

\textit{Complexity Analysis:}
The time complexity of Algorithm~\ref{Alg1} depends on the number of iterations, i.e., $\Gamma$, and the time complexity of one iteration.
In the $k$-th iteration, the $J$ individuals should first be sorted according to cost $o^{k,j}$, with a time complexity of $O(\log J)$, to support the selection of elite and parents individuals (\emph{lines} 3 and 5). 
To generate the $j$-th ($ 1 \leq j \leq J-1$) offspring individual, the crossover and mutation operations for generating candidate individual $\hat{\mathbf{V}}$ have $O(M)$ and $O(1)$ time complexity respectively, where $M$ is the number of genes (\emph{lines} 8-13).
To check the feasibility and calculate the cost of $\hat{\mathbf{V}}$, the transformed SOCP problem should be solved by using interior-point methods (\emph{line} 14), which requires at most a number of $O(\sqrt{2|\mathcal{L}^{\mathtt{A}}|})$ iterations to decrease the duality gap to a constant fraction of itself. The work for each iteration has a time complexity of $O(4|\mathcal{L}^{\mathtt{A}}|)$~\cite{lobo1998applications}. 
As there are at most $N\choose 2$ activated V2V links and $N$ activated V2I links for a total number of $N$ CAVs, we have $|\mathcal{L}^{\mathtt{A}}| \leq {N\choose 2} + N$, leading to a maximum total time complexity of $O(N^3)$ for solving the SOCP problem.  
The operations in \emph{lines 15-17} have a time complexity of $O(1)$. 
Hence, the time complexity for one iteration is $O(\log J) + (J-1)[O(M)p_C + O(1)p_M + O(N^3) + O(1)]$, which is equivalent to $O(J N^3)$ under the assumption of comparable object number $M$ and CAV number $N$, leading to a total time complexity of $O(\Gamma J N^3)$ for Algorithm~\ref{Alg1}.

\section{Performance evaluation}
\label{sc:sim}

\subsection{Simulation Setup} \label{sec:set}

For training the accuracy estimation learning model, we consider bounding box partition resolution $K\in\{1,2,3,4\}$. The default value of $K$ is set to $3$. 
Accordingly, the input dimension of the accuracy estimation learning model, i.e., $(K^3 + 3)$, takes candidate values from $\{4,11,30,67\}$. 
The learning model has two hidden DNN layers, with $(32,16)$ neurons and \texttt{Relu} activation functions between the input and output layers.  
For each $K$ value, to offline pre-train a corresponding accuracy estimation learning model, we first create a training dataset consisting of $5600$ labeled training data samples according to Subsection~\ref{sec:DNN-Based Classification Accuracy Prediction}. 
Specifically, we use an automated driving toolbox in MATLAB to simulate the random autonomous driving scenarios, 
and use a well-known VoxelNet model developed for 3D LiDAR point cloud processing as the object classification AI model~\cite{zhou2018voxelnet}.  
With the training dataset, we can train the accuracy estimation learning model by minimizing the mean square error (MSE) between the estimated object classification accuracy and the ground-truth labels.  

\begin{table}[t]
\small
\centering
\caption{\scshape{System parameters in simulation}}
\begin{tabular}{ p{0.72\columnwidth} | c} 
\hline\noalign{\vskip 0.3mm}\hline\noalign{\smallskip}
Parameters & Value \\
\noalign{\smallskip}\hline\noalign{\smallskip}
Bandwidth ($B$)&  $20$ MHz \tabularnewline
Noise power ($\sigma^2$) & $10^{-13}$ W \tabularnewline 
Transmit power ($P_n$) &  $1$ W \tabularnewline 
Path-loss exponent ($\gamma$) & 3.4 \tabularnewline
Channel fading coefficient ($|h_{n,n'}|^2$) & 1\tabularnewline 
Available computing resources at CAV $n$ ($f_n$) & $10$ GHz \tabularnewline
Available computing resources at RSU ($f_N$) & $200$ GHz \tabularnewline
Data size per observation point ($\varphi$) & $192$ bit \tabularnewline
\hline\noalign{\vskip 0.3mm}\hline\noalign{\smallskip}
\end{tabular}
\label{Table:System parameters in simulation}
\end{table}

For performance evaluation of the proposed cooperative sensing and computing scheme, we 
consider one ego CAV and three assisting CAVs, with a $360^\circ$ LiDAR sensor mounted on the roof of each CAV, on a $50$ $m$ unidirectional road segment in the coverage of one RSU, as illustrated in Fig.~\ref{fig:Network_scenario}.   
There are six objects being distributed in the ego CAV's RoI, which belong to four classes and include two trucks, two cars, one pedestrian and one cyclist.
We set the perception delay requirement as $T=20ms$, and set the accuracy requirement, $A$, among two candidate values in $\{0.7,0.9\}$. 
Other system parameters are given in Table~\ref{Table:System parameters in simulation}.
For simplicity, we assume identical received noise power, transmit power, channel fading coefficient, and computing capability among all the CAVs. 
The computation intensity (in cycle/point), $\epsilon$, is selected among four candidate values in $\{10000,20000,30000,40000\}$,  with 
$\epsilon=30000$ by default. 
We consider equal importance on minimizing the total computing and communication costs, by setting $\omega=0.5$. 
For the GA-based iterative solution, we set $p_C=0.9$ and $p_M=0.1$.

\subsection{Simulation Results} 
\label{sec:res}

\begin{figure}[!t]
\centering
\subfloat[]{\includegraphics[width=1.83in]{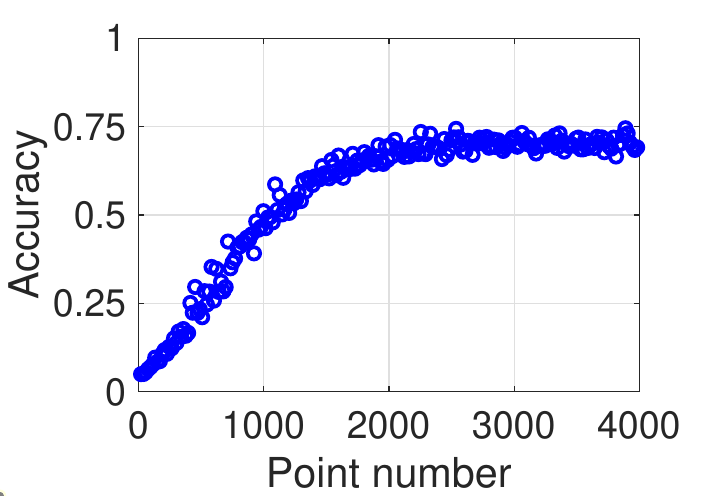}%
}
\subfloat[]{\includegraphics[width=1.83in]{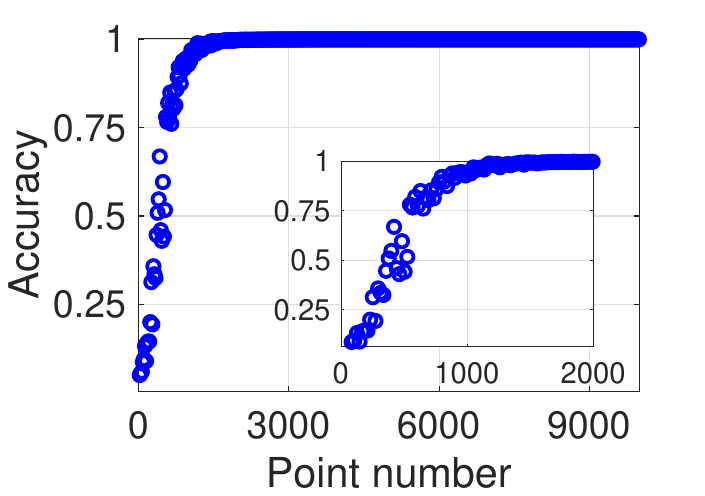}%
}
\caption{Object classification accuracy for object $0$. (a) Using data from ego CAV $0$. (b) Using data from assisting CAV $1$.}
\label{fig:div}
\end{figure}

\begin{figure}[!t]
\centering
\subfloat[]{\includegraphics[width=1.83in]{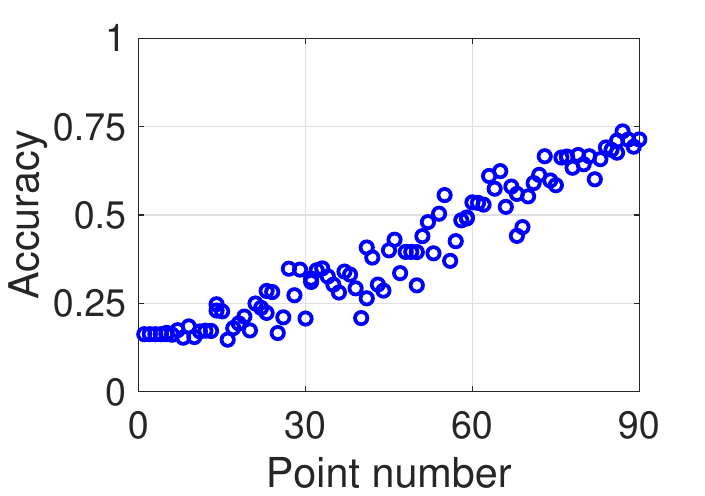}%
}
\subfloat[]{\includegraphics[width=1.83in]{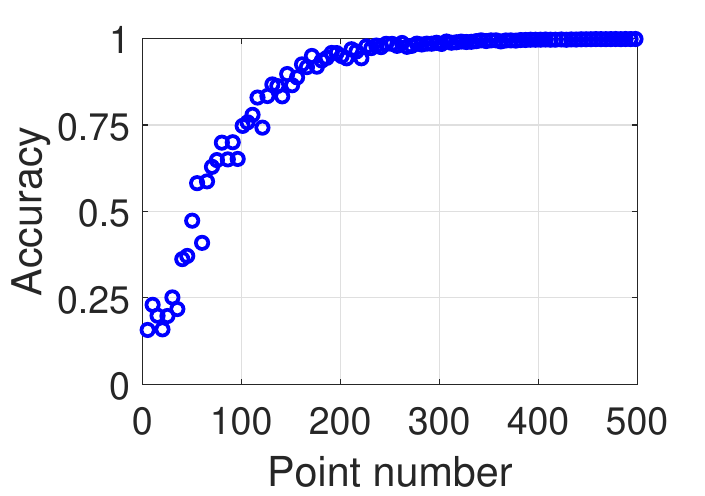}%
}
\caption{Object classification accuracy for object 2. (a) Using data from assisting CAV $1$. (b) Using data from assisting CAV $3$.}
\label{fig:bycdiv}
\end{figure}

We first demonstrate the necessity of including both the data quality indicator and the bounding box dimensions in the inputs of the accuracy estimation learning model. Specifically, we evaluate the impact of point number and spatial distribution of observation points in the object sensing data on the object classification accuracy for two example objects with different dimensions. 
We use object 0, a bigger-size truck, and object 2, a smaller-size cyclist, as examples. 
For the truck, we consider two sets of object sensing data from ego CAV $0$ and from assisting CAV $1$ respectively. 
Each set contains object sensing data with different resolutions, generated by down-sampling the corresponding original object sensing data using different down-sampling ratios from $0.01$ to $1$. 
For the cyclist, we consider two sets of object sensing data from assisting CAVs 1 and 3. 
Fig.~\ref{fig:div} and Fig.~\ref{fig:bycdiv} show the object classification accuracy achieved by using the corresponding two sets of object sensing data, for the truck and the cyclist, respectively. 

As the observation points of ego CAV $0$ for the truck are concentrated at the back side with a low sensing data diversity, we see in Fig.~\ref{fig:div}(a) that the object classification accuracy increases slowly to between $70\%$ and $75\%$ as the point number increases and gradually saturates with a further increase of point number, inferring that adding more observation points without improving the sensing data diversity
brings limited accuracy gain, 
when the point number is already large. 
By contrast, as the observation points of assisting CAV $1$ for the truck mainly spread over the front and left sides, providing more sensing data diversity, 
we see in Fig.~\ref{fig:div}(b) that the accuracy increases more rapidly with the increase of point number and approaches $100\%$ with less than $1500$ observation points, inferring that less observation points with better sensing data diversity are required to achieve the same accuracy. 
For the cyclist, we have similar observations in Fig.~\ref{fig:bycdiv} for the impact of point number and spatial distribution of object sensing data on the object classification accuracy.
By comparing Fig.~\ref{fig:div} and Fig.~\ref{fig:bycdiv}, we observe a significant impact of object size on object classification accuracy. 
For example, to satisfy an accuracy requirement of $A = 0.9$, $1000$ points are required from assisting CAV $1$ for the truck, while $200$ points are required from assisting CAV $3$ for the cyclist. 
Therefore, it is necessary to consider both the data quality indicator, which characterizes the volume and spatial distribution of object sensing data, and the bounding box dimensions in learning the object classification accuracy estimation function. 

\begin{table}[t]
\small
\centering
\caption{\scshape{MSE of Accuracy Estimation Learning Model for Different Partition Resolution ($K$)}}
\begin{tabular}{ c | c| c| c| c} 
\hline\noalign{\vskip 0.3mm}\hline\noalign{\smallskip}
Partition resolution ($K$) & $1$ & $2$ & $3$ & $4$\\
\noalign{\smallskip}\hline\noalign{\smallskip}
MSE &  $0.171$ & $0.059$ & $0.049$ & $0.046$ \tabularnewline
\hline\noalign{\vskip 0.3mm}\hline\noalign{\smallskip}
\end{tabular}
\label{Table:resolution}
\end{table}

\begin{table}[t]
\small
\centering
\caption{\scshape{Performance of Accuracy Estimation Learning Model for Different Object Types}}
\begin{tabular}{ c | c| c| c| c} 
\hline\noalign{\vskip 0.3mm}\hline\noalign{\smallskip}
Metric & Car & Truck & Pedestrian & Cyclist\\
\noalign{\smallskip}\hline\noalign{\smallskip}
MSE &  $0.095$ & $0.043$ & $0.029$ & $0.011$ \tabularnewline
MAE &  $0.163$ & $0.100$ & $0.135$ & $0.068$ \tabularnewline
VAE &  $0.068$ & $0.033$ & $0.011$ & $0.006$ \tabularnewline
\hline\noalign{\vskip 0.3mm}\hline\noalign{\smallskip}
\end{tabular}
\label{Table:object}
\end{table}

For each candidate value of bounding box partition resolution $K\in\{1,2,3,4\}$, an accuracy estimation learning model with a $K^3+3$ input dimension is trained based on a corresponding training dataset.   
Table~\ref{Table:resolution} shows the relationship between $K$ and the MSE between the estimated object classification accuracy and the ground-truth labels. 
We observe that an accuracy estimation learning model corresponding to a larger $K$ value has a smaller MSE, indicating a higher accuracy in estimating the object classification accuracy. 
The reason is that the data quality indicator captures the spatial distribution of observation points in more details for a larger $K$ value. 
For example, when $K$ is equal to $1$, the data quality indicator is reduced to a scalar representing the total point number, losing the spatial distribution information. 
When $K$ is increased to $2$, the data quality indicator has $K^3=8$ dimensions, which can coarsely capture the spatial distribution of observation points. Hence, a big drop is observed in the MSE. 
However, the improvement in MSE becomes less significant by further increasing $K$.
The potential reason is that there are too many zeros in the data quality indicator when $K$ is large, which does not contribute to significantly more information gain in characterizing the data spatial distribution, especially when the observation points for an object are very concentrated. 

We also evaluate the performance of the accuracy estimation learning model for different object types at a same partition resolution, $K=3$. 
Let $\varepsilon_i$ denote the error between the estimated accuracy by the model and the label for the $i$-th data sample in a training dataset of size $I$. 
Three metrics are considered, including the MSE, $\frac{1}{I} \sum_{i=1}^I \varepsilon_i^2$, the mean absolute error (MAE), $\bar{\varepsilon} = \frac{1}{I} \sum_{i=1}^I \left|\varepsilon_i\right|^2$, and the variance of absolute error (VAE), $\frac{1}{I} \sum_{i=1}^I \left(|\varepsilon_i| - \bar{\varepsilon}\right)^2$. 
The model performance for car, truck, pedestrian, and cyclist are summarized in Table~\ref{Table:object}. 
It is observed that the accuracy estimation has disparate performance among different object types. However, with the consideration of the small MSE values below $0.1$, we conclude that the model has good performance for all object types despite the slight differences.

\begin{figure}[!t]
\centering
\subfloat[]{\includegraphics[width=3.0in]{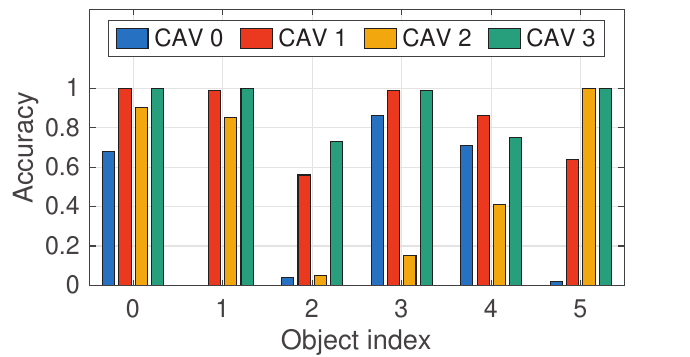}%
}
\hfil
\subfloat[]{\includegraphics[width=3.0in]{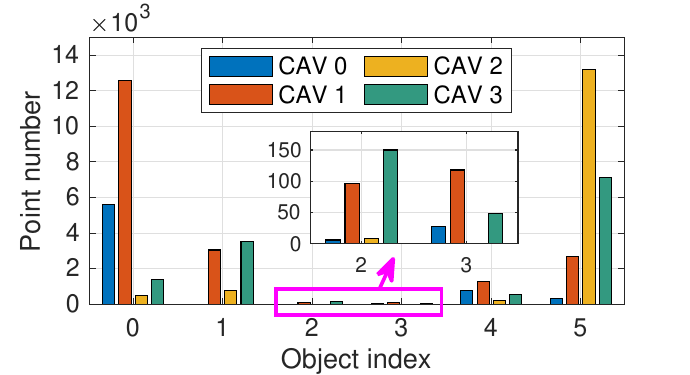}%
}
\caption{Object classification performance without cooperative sensing. (a) Object classification accuracy. (b) Point number.}
\label{fig:single_AV_perf}
\end{figure}

To demonstrate the benefit of cooperative sensing in term of accuracy improvement, we first examine the object classification accuracy without cooperative sensing for each object, with results shown in Fig.~\ref{fig:single_AV_perf}(a). For a better illustration, Fig.~\ref{fig:single_AV_perf}(b) shows the point number in the object sensing data at each CAV for each object for reference. 
Fig.~\ref{fig:single_AV_perf}(a) shows that none of the CAVs can achieve an accuracy beyond $0.7$ for all the objects by purely relying on its own sensing data, demonstrating the necessity for cooperative sensing under the simulation settings.
The poor accuracy below 0.7 of different CAV-object pairs is due to different reasons, such as occlusion, distance and limited viewing angle, which can be observed from the spatial relationships among CAVs and objects in Fig.~\ref{fig:Network_scenario}. 
We also observe that less observation points are required to achieve a similar accuracy for objects with a smaller size. For example, CAV 1 achieves an accuracy close to 1 for object 1 (a large-size car) and object 3 (a small-size pedestrian) by using around 3000 and less than 150 observation points, respectively.

\begin{figure*}[t]
\centering
\includegraphics[width=0.85\linewidth]{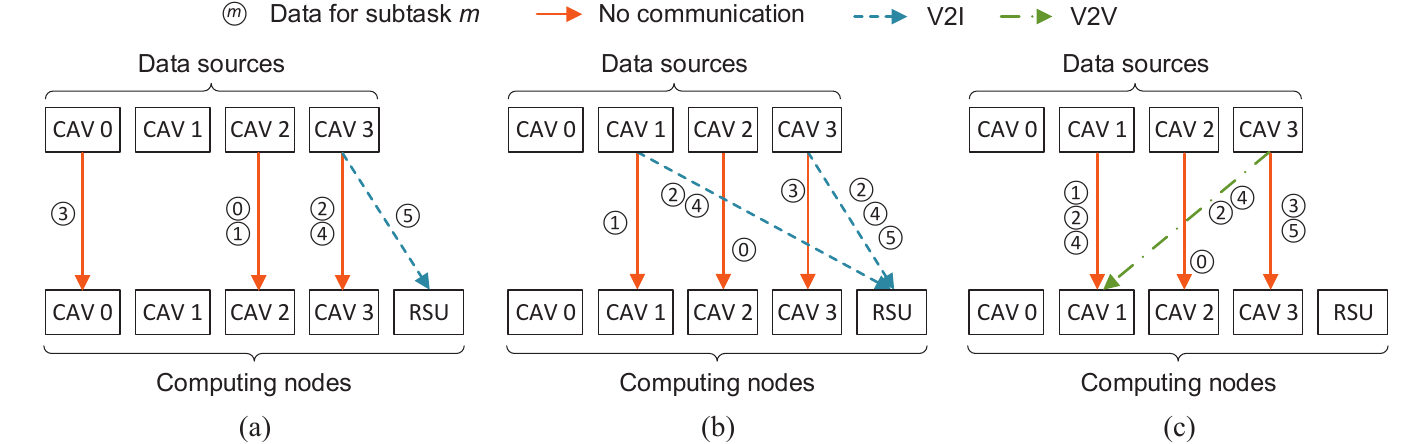}
\caption{Data selection and subtask placement solutions under different conditions. (a) {$A=0.7$}, $\epsilon=30000,40000$. (b) {$A=0.9$}, $\epsilon=30000,40000$. (c) {$A=0.9$}, $\epsilon=10000,20000$.} 
    \label{fig:cooperation_accuracy}
\end{figure*}

Next, we evaluate the performance of the proposed accuracy-aware cooperative sensing and computing scheme, based on a trained accuracy estimation learning model with $K=3$. Fig.~\ref{fig:cooperation_accuracy} illustrates how the data selection and subtask placement strategies change among subtasks in the simulated scenario based on the proposed scheme under different conditions.  
Each arrow in the figure indicates a sensing data flow from a data source to a computing node. 
For example, an arrow starting from CAV $n$ to computing node $n'$ is denoted by $(n,n')$. 
The numbers in circles beside arrow $(n,n')$ indicate the indexes of subtasks with sensing data selection from CAV $n$ and placement at computing node $n'$. 
If $n=n'$, no communication is required for the selected sensing data, as the sensing data are processed at the CAV which holds the data.
If $n\neq n'$, the arrow indicates V2V communication for $n'<N$ or V2I communication for $n'=N$, with $N=4$ in the simulated scenario.

Fig.~\ref{fig:cooperation_accuracy}(a) and Fig.~\ref{fig:cooperation_accuracy}(b) show the data selection and subtask placement strategies for different accuracy requirements at high computation intensity $\epsilon=30000,40000$. 
For a lower accuracy requirement at $A=0.7$, no cooperative sensing is required, and each subtask just uses the object sensing data from a single CAV, as illustrated in Fig.~\ref{fig:cooperation_accuracy}(a). Most subtasks except subtask 5 consume no communication resources, by processing the selected sensing data locally. 
Each of them is placed at a CAV which holds the minimum amount of object sensing data that can satisfy the accuracy requirement, $A=0.7$, for the corresponding object. 
For subtask 5, CAV 3 has the minimum sensing data with accuracy satisfaction, but it cannot process the data locally with delay satisfaction due to the large data size. 
Hence, subtask 5 is offloaded to the RSU with more computing resources for delay improvement, which is supported by V2I communication. 
However, when the accuracy requirement becomes more stringent for $A=0.9$, cooperative sensing is required for subtasks $2$ and $4$, as illustrated in Fig.~\ref{fig:cooperation_accuracy}(b). 
Specifically, CAVs 1 and 3 are selected to provide sensing data for both subtasks $2$ and $4$. 
Due to the half-duplex communication constraints, subtasks $2$ and $4$ that use sensing data from both CAVs 1 and 3 must be placed at the same computing node, which can be CAV 1, CAV 3, or the RSU. 
As CAV 1 and CAV 3 are the preferred data selection and computing nodes for subtask 1 and subtask 3 respectively, neither of them can support the extra computation for both subtasks 2 and 4 with delay satisfaction due to the limited local computing resources. 
Then, the RSU is the only feasible computing node for subtasks 2 and 4.

\begin{figure}[!t]
\centering
\subfloat[]{\includegraphics[width=1.77in]{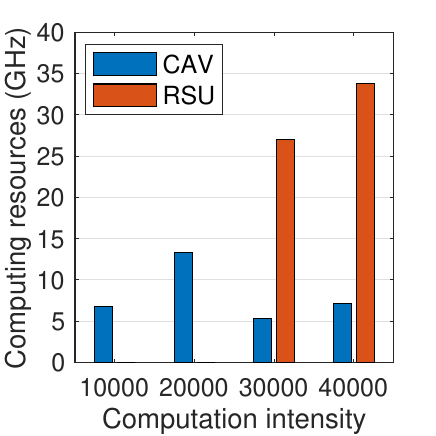}%
}
\subfloat[]{\includegraphics[width=1.77in]{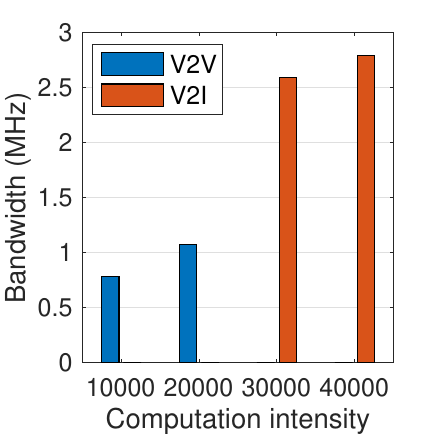}%
}
\caption{Resource consumption versus computation intensity $\epsilon$ for $A=0.9$. (a) Computing resources. (b) Communication resources.}
\label{fig:ave}
\end{figure}

Fig.~\ref{fig:cooperation_accuracy}(b) and Fig.~\ref{fig:cooperation_accuracy}(c) show the data selection and subtask placement strategies for different computing intensities ($\epsilon$) at a high accuracy requirement, $A=0.9$. 
The corresponding total computing and communication resource consumptions for different $\epsilon$ values are shown in Fig.~\ref{fig:ave}. 
We see from Fig.~\ref{fig:ave}(a) that the total computing resource consumption at the RSU and CAVs increases almost linearly with $\epsilon$, as the computing resource demand for each subtask is in proportion to the computation intensity. 
By comparing Fig.~\ref{fig:cooperation_accuracy}(b) and Fig.~\ref{fig:cooperation_accuracy}(c), we see that the accuracy-aware data selection strategies for the subtasks keep unchanged with the increase of $\epsilon$, indicating a constant total selected sensing data size, but the subtask placement solutions are adjusted as $\epsilon$ increases from $20000$ to $30000$, to accommodate the proportionally increasing total computing demand with $\epsilon$.   
With low computation intensities at $\epsilon=10000$ and $20000$, all the subtasks are placed at CAVs for communication resource efficiency, as illustrated in Fig.~\ref{fig:cooperation_accuracy}(c). The selected sensing data from CAVs are either locally processed without data transmission or processed at other CAVs via V2V-based data transmission. 
Accordingly, there is no RSU computing resource consumption and no V2I communication resource consumption, as shown in Fig.~\ref{fig:ave} for $\epsilon=10000$ and $20000$. 
In comparison, with high computation intensities at $\epsilon=30000$ and $40000$, there are computing resource consumptions at both CAVs and RSU, as subtasks 2, 4, and 5 are offloaded to the RSU for delay satisfaction, as illustrated in Fig.~\ref{fig:cooperation_accuracy}(b) and Fig.~\ref{fig:ave}(a). 
The selected sensing data are transmitted to the RSU via V2I links, requiring more transmission bandwidth for delay satisfaction due to more data and longer distances for transmission in comparison with V2V links, as shown in Fig.~\ref{fig:ave}(b). 
We also observe from Fig.~\ref{fig:ave}(b) that, when the joint data selection and subtask placement solution keeps unchanged while $\epsilon$ increases, e.g., from $10000$ to $20000$, or from $30000$ to $40000$, the bandwidth consumption for transmitting the same amount of selected sensing data experiences a slight increase to reduce the transmission delay, which compensates for the higher computing delay at a higher computing intensity.

\begin{table}[t]
\footnotesize
\centering
\caption{\scshape{Summary of Benchmark and Proposed Schemes}}
\label{Table:benchmarks}
\begin{tabular}{ |c|c|c|c|c| }  
\hline
\multirow{3}{*}{\backslashbox{\textbf{Scheme}}{\textbf{Feature}}} & \multicolumn{2}{c|}{ Data Selection } & \multicolumn{2}{c|}{  Computing }\\ \cline{2-5}
& \multirow{2}{*}{Granularity} & Accuracy & \multirow{2}{*}{RSU} & \multirow{2}{*}{CAV} \\ 
& &  Awareness & & \\\hline
\textsl{All} & Full &   x & \checkmark & x\\ \hline 
\textsl{Unified} & Full &  \checkmark & \checkmark & x\\ \hline
\textsl{Nearest} & Object &   x & \checkmark & \checkmark \\ \hline
\textsl{Centralized} & Object &  \checkmark & \checkmark &  x \\ \hline
\textsl{Proposed} & Object &  \checkmark & \checkmark & \checkmark \\ \hline
\end{tabular}
\end{table}

Moreover, we compare the performance between the proposed and four benchmark solutions for raw-level cooperative sensing. 
Table~\ref{Table:benchmarks} summarizes the key features of the schemes, in terms of the granularity and accuracy awareness for data selection and the computing nodes for (sub)task placement. 
In an \emph{all} scheme, all CAVs share the full raw sensing data, and the data fusion and processing are performed at an RSU. 
In a \emph{unified} benchmark, a most resource-efficient subset of CAVs are selected to share their full raw sensing data for data fusion and processing at the RSU, with accuracy satisfaction for all objects. 
Different from the proposed scheme where the data for each object can be selected from a different CAV group, such a scheme restricts the data selection from a unified CAV group for different objects, leading to potential accuracy over-provisioning for some objects. 
In a \emph{nearest} benchmark, data selection in per-object granularity is considered, but is limited to that of a nearest CAV for each object based on proximity principle. No accuracy awareness is explicitly considered. Each subtask is placed at either the nearest CAV or the RSU, depending on whether the local computing capability can support the subtask with delay satisfaction or not. 
In a \emph{centralized} benchmark, the data selection strategy is the same as that of the proposed scheme with per-object granularity and accuracy-awareness, but the difference lies in the centralized data fusion and processing at the RSU. 

\begin{figure}[!t]
\centering
\subfloat[]{\includegraphics[width=3in]{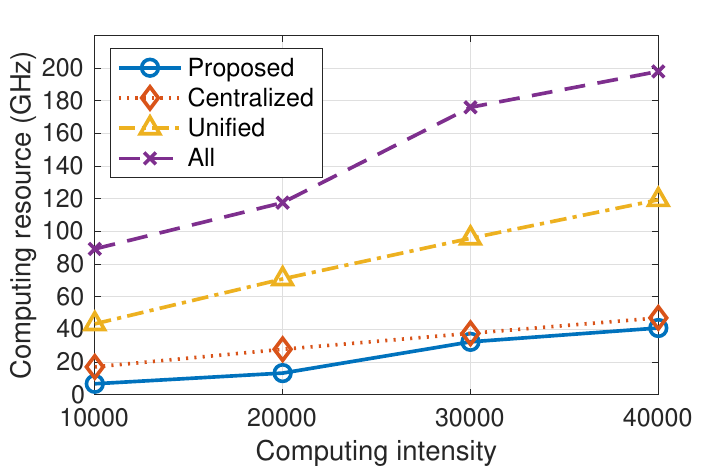}%
}
\hfil
\subfloat[]{\includegraphics[width=3in]{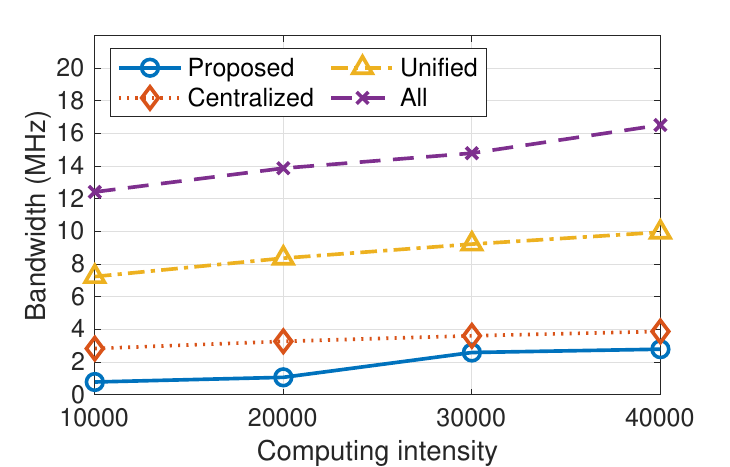}%
}
\caption{Performance comparison between the proposed and benchmark solutions as computation intensity $\epsilon$ increases for $A=0.9$. (a) Computing resources. (b) Communication resources.}
\label{fig:cd}
\end{figure}

\begin{table}[t]
\footnotesize
\centering
\caption{\scshape{Subtask Accuracy in Benchmark and Proposed Schemes}}
\label{Table:accuracy-comp}
\begin{tabular}{ |c|c|c|c|c|c|c| }  
\hline
\backslashbox{\textbf{Scheme}}{\textbf{Subtask}} & 0 & 1 & 2 & 3 & 4 & 5\\ \hline
\textsl{All} & $1$ & $1$ & $0.93$ & $1$ & $0.99$ & $1$\\ \hline 
\textsl{Unified} & $1$ & $1$ & $0.92$ & $1$ & $0.97$ & $1$\\ \hline
\textsl{Centralized} & $0.90$ & $0.99$ & $0.91$ & $0.99$ & $0.96$ & $1$ \\ \hline
\textsl{Proposed} & $0.90$ & $0.99$ & $0.91$ & $0.99$ & $0.96$ & $1$\\ \hline
\end{tabular}
\end{table}

Table~\ref{Table:accuracy-comp} summarizes the achieved accuracy for each subtask by using different schemes. 
Among all the solutions, only the \textit{nearest} benchmark cannot satisfy an accuracy requirement of $A=0.9$ for all the objects. 
Specifically, the accuracy requirement cannot be satisfied without data fusion for objects 2 and 4, and the nearest CAV 0 for object 0 provides limited sensing data diversity due to the close distance from the back side. 
Fig.~\ref{fig:cd} shows the resource consumption comparison between the proposed and the remaining three benchmark solutions, all of which satisfy both the accuracy and delay requirements, as computation intensity $\epsilon$ increases from $10000$ to $40000$ for $A=0.9$. 
We observe increasing communication and computing resource consumption as $\epsilon$ increases for all the solutions.
Specifically, due to the lack of accuracy-aware data selection, the \emph{all} scheme incurs the highest network resource cost for both computing and communication, 
with accuracy over-provisioning for all objects. 
In the \emph{unified} benchmark solution, CAV 1 and CAV 3 are selected to provide sensing data for all the subtasks with accuracy satisfaction, which significantly reduces the total selected sensing data and the resulting network resource consumption in comparison with the \emph{all} benchmark. 
For the \emph{centralized} and proposed solutions, the least amount of data are selected from different CAVs for accuracy satisfaction, due to the fine-grained accuracy-aware data selection in per-object granularity. 
Further, as the distributed computing among CAVs and RSU in the proposed solution reduces the amount of selected sensing data for transmission, by allowing selecting and processing sensing data at the same computing node, we observe a reduction in both the computing and communication resource usage in the proposed solution for delay satisfaction. 
From Table~\ref{Table:accuracy-comp}, we observe that the accuracy over-provisioning is not fully eliminated by the most resource-efficient proposed solution among all the solutions, due to the considered per-object data selection granularity. The proposed solution can be further improved by allowing data resolution reduction for the selected object sensing data, which remains as our future work.

\section{Conclusion}\label{sc:con}

In this paper, an accuracy-aware cooperative sensing and computing scheme is proposed for edge-assisted CAVs, based on a supervised learning model for accuracy estimation.  
By exploiting the parallelism among object classification subtasks, our proposed scheme facilitates fined-grained sensing data selection from the full raw sensing data of CAVs and allows distributed computation among CAVs and the RSU, which enhances the overall network resource efficiency while satisfying the delay and accuracy requirements. 
Simulation results demonstrate the effectiveness of the learning model in accuracy estimation.  
Further, the proposed scheme achieves accuracy awareness and 
resource efficiency in comparison with benchmark solutions.
In our future work, we will explore the data resolution configuration beyond the selection of sensing data, which has the potential to further enhance the resource efficiency. 
Moreover, as the communication network infrastructures are gradually embedded with sensing capability, the cooperative sensing between CAVs and RSUs will also be studied.

\bibliographystyle{IEEEtran}
\bibliography{ref}

\end{document}